\documentclass{aa}
\usepackage{lscape}
\usepackage{graphicx}
\usepackage{txfonts}
\usepackage{longtable}
\usepackage{natbib}[1]
\usepackage{url}
\usepackage{float}
\usepackage{aalongtable}
\usepackage{longtable}
\usepackage{textcomp}
\begin{document}

\title{The CARMENES search for exoplanets around M dwarfs}  
\subtitle{Photospheric parameters of target stars from high-resolution spectroscopy}
\titlerunning{Photospheric parameters of target stars}
\author{V.~M. Passegger\inst{1,2}
\and
A. Reiners\inst{1}
\and
S.~V. Jeffers\inst{1}
\and
         S. Wende-von Berg\inst{1}
         \and 
         P. Sch\"ofer\inst{1}
         \and
         J.~A. Caballero\inst{3,4}
         \and
         A. Schweitzer\inst{2}
         \and
         P.~J. Amado\inst{5}
         \and
         V. J. S. B\'ejar\inst{6}
         \and
         M. Cort\'es-Contreras\inst{3,7}
         \and
         A.P. Hatzes\inst{8}
         \and
         M. K\"urster\inst{9}
         \and
         D. Montes\inst{7}
         \and 
         S. Pedraz\inst{10}
         \and
         \newline A. Quirrenbach\inst{4}
         \and
         I. Ribas\inst{11}
         \and 
         W. Seifert\inst{4} 
         }
\institute{
        Institut f\"ur Astrophysik, Georg-August-Universit\"at, Friedrich-Hund-Platz 1, D-37077 G\"ottingen, Germany
        \and
        Hamburger Sternwarte, Gojenbergsweg 112, D-21029 Hamburg, Germany, 
        \email{vpassegger@hs.uni-hamburg.de}
        \and
         Departamento de Astrof\'isica, Centro de Astrobiolog\'ia (CSIC-INTA), Camino Bajo del Castillo s/n, ESAC Campus, E-28691 Villanueva de la Ca\~nada, Madrid, Spain
        \and
        Zentrum f\"ur Astronomie der Universt\"at Heidelberg, Landessternwarte, K\"onigstuhl 12, D-69117 Heidelberg, Germany
        \and
         Instituto de Astrof\'isica de Andaluc\'ia (IAA-CSIC), Glorieta de la Astronom\'ia s/n, E-18008 Granada, Spain 
        \and
        Instituto de Astrof\'{\i}sica de Canarias, V\'ia L\'actea s/n, E-38205 La Laguna, Tenerife, Spain, and Departamento de Astrof\'{\i}sica, Universidad de 
         La Laguna, E-38206 La Laguna, Tenerife, Spain
         \and
         Departamento de Astrof\'{\i}sica y Ciencias de la Atm\'osfera, Facultad de Ciencias F\'{\i}sicas, Universidad Complutense de Madrid, E-28040 Madrid, Spain
        \and
         Th\"uringer Landessternwarte Tautenburg, Sternwarte 5, D-07778 Tautenburg, Germany
        \and
        Max-Planck-Institut f\"ur Astronomie, K\"onigstuhl 17, D-69117 Heidelberg, Germany
        \and
         Centro Astron\'omico Hispano-Alem\'an (CSIC-MPG), Observatorio Astron\'omico  de  Calar  Alto,  Sierra  de  los  Filabres, E-04550  G\'ergal, Almer\'ia, Spain
         \and
         Institut de Ci\`encies de l'Espai (CSIC-IEEC), Can Magrans s/n, Campus UAB, E-08193 Bellaterra, Barcelona, Spain
         }

\date{}

\abstract
{The new CARMENES instrument comprises two high-resolution and high-stability spectrographs that are used to search for habitable planets around M dwarfs in the visible and near-infrared regime 
via the Doppler technique.} 
{Characterising our target sample is important for constraining the physical properties of any planetary systems that are 
detected. The aim of this paper is to determine the fundamental stellar parameters of the CARMENES M-dwarf target sample from high-resolution spectra observed with CARMENES. We also include 
several M-dwarf spectra observed with other high-resolution spectrographs, that is CAFE, FEROS, and HRS, for completeness. } 
{We used a $\chi^2$ method to derive the stellar parameters effective temperature $T_{\rm eff}$, surface gravity $\log{g}$, and metallicity [Fe/H] of the target stars by fitting the most 
recent version of the PHOENIX-ACES models to high-resolution spectroscopic data. These stellar atmosphere models incorporate a new equation of state to describe spectral features of 
low-temperature stellar atmospheres. Since $T_{\rm eff}$, $\log{g}$, and [Fe/H] show degeneracies, the surface gravity is determined independently using stellar evolutionary models.}
{We derive the stellar parameters for a total of 300 stars. The fits achieve very good agreement between the PHOENIX models and observed spectra. We estimate that our method provides parameters with 
uncertainties of $\sigma_{T_{\rm eff}} = 51$\,K, $\sigma_{\log{g}} = 0.07$, and $\sigma_{\rm [Fe/H]} = 0.16$, and show that atmosphere models for low-mass stars 
have significantly improved in the last years. Our work also provides an independent test of the new PHOENIX-ACES models, and a comparison for other methods using low-resolution spectra. 
In particular, our effective temperatures agree well with literature values, while metallicities determined with our method exhibit a larger spread when compared to literature results.}
{}
 \maketitle
%

\section{Introduction}

M dwarfs are of great interest for current exoplanet searches. Compared to Sun-like stars, M dwarfs have lower stellar masses and smaller radii, which facilitates detecting orbiting 
planets, especially those within the habitable zone (i.e. the orbital distance from the star at which liquid water can exist on the surface of the planet). Within this context, the 
Calar Alto high-Resolution search for M dwarfs with Exo-earths with Near-infrared and optical \'Echelle Spectrographs (CARMENES) instrument was built to search for rocky planets 
in the habitable zones of M dwarfs via the Doppler technique. 
CARMENES is mounted on the Zeiss 3.5\,m telescope at Calar Alto Observatory, located in Almer\'{\i}a, in southern Spain. After commissioning at the end of 2015 
\citep[see][]{Quirrenbach2016}, CARMENES has been taking data since January 1, 2016. 
The instrument consists of two fiber-fed spectrographs spanning the visible and near-infrared wavelength range, from 0.52 to 0.96\,$\mu$m and from 0.96 to 1.71\,$\mu$m, with a  
spectral resolution of R $\approx$ 94,600 and 80,500, respectively. Simultaneous observations in two wavelength ranges are favourable for distinguishing between a planetary signal 
and stellar activity, which can mimic a false-positive signal. Both spectrographs are designed to perform high-accuracy
radial-velocity measurements with a long-term stability of $\sim$1\,m\,s$^{-1}$ \citep[][]{Quirrenbach2014,Reiners2017}, with the aim of being able to detect 2 M$_{\oplus}$ 
planets orbiting in the habitable zone of M5\,V stars. 

To select the most promising targets, an extensive literature search was carried out \citep{Alonso2015,Caballero2016a}. Additional observations were conducted with low- 
and high-resolution spectrographs and high-resolution imaging. A first paper about the CARMENES science preparation was published by \cite{Alonso2015}. They focused on the 
determination of spectral types and activity indices from low-resolution spectra and also gave a description of the CARMENES target sample. \cite{Cortes2017} searched for close 
low-mass companions in the CARMENES target sample and analysed possible multiplicity using lucky imaging data. \cite{Jeffers2018} determined rotational velocities 
and H$\alpha$ activity indices measured from high-resolution spectra taken with CAFE and FEROS.  The Carmencita database 
\citep[CARMENes Cool dwarf Information and daTa Archive,][]{Alonso2015} contains all the 
information collected from the target sample, that is, astrometry; distances; spectral types; photometry in 20 different bands; X-ray count rates and hardness ratios; H$\alpha$ emission; 
rotational, radial, and Galactocentric velocities; stellar and planetary companionship; membership in open clusters and young moving groups; and targets in other radial-velocity surveys.

Because of their lower temperatures, M dwarfs show more complex spectra than Sun-like stars. Forests of spectral features caused by molecular lines make the determination of atmospheric 
parameters more difficult and require a full spectral synthesis. This necessitates the use of accurate atmosphere models that reproduce the spectral features present in cool star 
spectra. The PHOENIX-ACES models that we used here were presented by \cite{Husser2013}. 

It is important for planet search surveys to determine fundamental stellar parameters to be able to characterise the system. 
\citet[hereafter GM14]{GaidosMann2014} observed $JHK$-band spectra of 121 M dwarfs. About half of them were also 
observed in the visible range. The authors determined effective temperatures in the visible by fitting BT-Settl models \citep{Allard2012a} to their spectra. For stars without spectra in 
the visible, they calculated spectral curvature indices from $K$-band spectra to determine effective temperatures. They derived metallicities using the relation of the 
atomic line strength in the visible, $J, H,$ and $K$ bands as defined in \cite{Mann2013}. The relations were calibrated using binaries with F, G, and K primary stars that have an 
M-dwarf companion. 
The BT-Settl models were also used by \citet[hereafter RA12]{RojasAyala2012}, who determined temperatures and metallicities of 133 M dwarfs in the near-infrared $K$ band with 
mid-resolution TripleSpec spectra ($R \sim$ 2700). They measured the equivalent widths of Na~{\sc i} and Ca~{\sc i} and the H$_2$O-K2 index, quantifying the absorption due to 
H$_2$O opacity by using BT-Settl models \citep{Allard2012a} with solar metallicity. 

\cite{Rajpurohit2013} also used the models by \cite{Allard2012b} to calculate effective temperatures for 152 M dwarfs with low- and mid-resolution spectra. They found that the overall 
slope of model and observed spectra matched very well, although there were still some discrepancies in the depth of single lines and absorption bands. 

Another widely used set of models 
are the MARCS models \citep{Gustafsson2008}. Among others, \cite{LindgrenHeiter2017} used these models together with the package Spectroscopy Made Easy \citep[SME --][]{ValentiPiskunov1996} 
to determine metallicities for several M dwarfs from fitting several atomic species in the near-infrared. \cite{Souto2017} also fitted MARCS models to high-resolution APOGEE spectra 
to derive abundances for 13 elements of the exoplanet-hosting M dwarfs Kepler-138 and Kepler-186. \cite{Veyette2017} combined spectral synthesis, empirical calibrations, and 
equivalent widths to derive precise temperatures as well as Ti and Fe abundances from high-resolution M-dwarf spectra in the near-infrared. 
A more detailed overview of the different approaches on the determination of stellar parameters can be found in \cite{Passegger2016}. In contrast to the above mentioned works, we here 
analyse a large sample of 300 M dwarfs by fitting high-resolution spectra to the most advanced model spectra using broad wavelength ranges. 
We obtain $T_{\rm eff}$, $\log{g}$, and [Fe/H] for all target stars from spectra taken with CARMENES, FEROS, CAFE, and HRS, compare our results with the literature, and show our 
conclusions. 

\begin{figure}[]
  \includegraphics[width=0.45\textwidth,bb= 0 0 400 410]{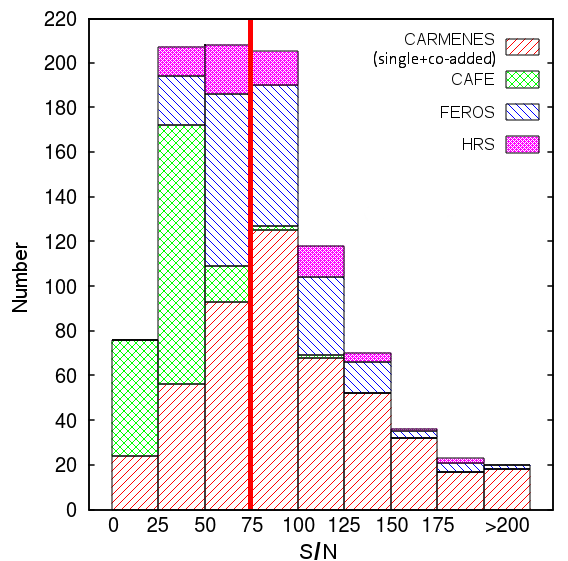}
  \caption{Histogram distribution of the signal-to-noise ratios for all spectra from all four spectrographs. The red solid line marks the signal-to-noise ratio limit of 75. }
  \label{fig:S/N_histo}
  \end{figure}


\section{Observations}
\label{Observations}
We obtained 973 spectra of 544 stars with spectral types between M0.0\,V and M8.0\,V with CARMENES and the high-resolution spectrographs CAFE, FEROS, and HRS. 
The Calar Alto Fiber-fed Echelle spectrograph (CAFE) is mounted at the 2.2\,m telescope of the Calar Alto Observatory in Spain \citep{Aceituno2013}. 
The Fiber-fed Extended Range Optical Spectrograph (FEROS) spectrograph is an echelle spectrograph located at the 2.2\,m telescope at the ESO La Silla Observatory in Chile 
\citep{Kaufer1997,Stahl1999}. 
The High Resolution Spectrograph (HRS) is an echelle spectrograph mounted at the 9.2\,m Hobby-Eberly telescope at McDonald Observatory in Texas, USA \citep{Tull1998}. 
For a detailed description of the observations and the reduction process of CAFE, FEROS, and HRS data, we refer to \cite{Jeffers2018}. The properties of the 
spectrographs and observations are summarised in Table~\ref{tab:obs}. 
The CARMENES spectra were reduced automatically every night by the CARMENES pipeline \citep{Caballero2016b}. In our analysis we also used the co-added CARMENES spectra, which 
are produced by the SERVAL pipeline to measure radial-velocity shifts \citep{Zechmeister2017, Reiners2017}. For each star, the co-added spectrum consists of at least five single observations 
that are co-added to increase the signal-to-noise ratio (S/N). 

We found that for most spectra with S/N<75, the temperatures and metallicities were either unrealistically high or low, therefore we set a general S/N limit of 75 for all spectra. 
In order to examine spectra with the highest S/N, we first analysed all co-added CARMENES spectra, followed by single CARMENES spectra for stars without co-added spectra. 
We also investigated stars that are not being monitored by CARMENES for completeness, therefore we included spectra from FEROS, CAFE, and HRS in our analysis. 
When the same star was observed with more than one instrument, we selected the observation with higher S/N. \cite{Passegger2017} showed that parameters derived from spectra from 
different spectrographs are comparable with deviations smaller than the typical uncertainty for these parameters. A histogram distribution showing the S/Ns for all spectra is presented 
in Fig.~\ref{fig:S/N_histo}. After applying the S/N limit we finally determined parameters of 300 different M dwarfs, 235 of which were observed with CARMENES.

\begin{center}
  \begin{table*}
    \caption{Summary of observations and analysed stars.}
    \label{tab:obs}
    \centering %
    \begin{tabular}{lcccccc}
      \hline \hline 
      Spectrograph & Resolution & $\Delta \lambda$ [nm] & Number of & Number of & Number of & Observing period\\ 
		   &		&		& spectra (observed)   & stars (observed) & stars (results) & 	\\
	   \hline 
      CARMENES & \textasciitilde 94600 & 550-1700 & 485 & 338 & 235 & 2016-01-01 to 2017-06-30\\
      CAFE  & \textasciitilde 62000 & 396-950  & 187& 77 & 2 &  2013-01-21 to 2014-09-26\\
      FEROS & 48000 & 350-920 & 222 & 107 & 55 & 2012-12-31 to 2014-07-11\\
      HRS   & 60000 & 420-1100 & 79 & 22& 8 & 2011-09-29 to 2013-06-18\\
	  \hline
      Total & ... & ... & 973 & 544 & 300 & ...\\
    
      \hline
    \end{tabular}
  \end{table*}
\end{center}

  \begin{center}
  \begin{table*}
    \caption{Wavelength regions and lines used for the $\chi^2$ fitting.}
    \label{tab:lines}
    \centering %
    \begin{tabular}{l|ccccc}
      \hline \hline 
      Line/band & $\gamma$-TiO &  K~{\sc i} & Ti~{\sc i} & Fe~{\sc i} & Mg~{\sc i} \\ 
	   \hline 
	$\lambda_c$[nm] & 705.5 & 770.1 & 841.5, 842.9, 843.7, 843.8&  847.1, 851.6, 867.7 & 880.9\\
	  		&  &  & 846.9, 867.8, 868.5 & 869.1, 882.7&  \\

	  \hline
    
      \hline
    \end{tabular}
  \end{table*}
\end{center}


\section{Method}
\label{Method}
We adapted the method described in \cite{Passegger2016}, who determined the fundamental stellar parameters effective temperature $T_{\rm eff}$, surface gravity $\log{g}$, 
and metallicity [Fe/H] for four M dwarfs using the latest grid of PHOENIX model spectra presented by \citet{Husser2013}. 
The PHOENIX code was developed by \cite{Hauschildt1992,Hauschildt1993} and has been considerably improved since then (e.g. \citealt{Hauschildt1997}; \citealt{HauschildtBaron1999};
\citealt{Claret2012}; \citealt{Husser2013}). The code can generate 1D model atmospheres of plane-parallel or spherically symmetric stars and degenerate objects (late-type stars as well as brown dwarfs,
white dwarfs, and giants), accretion discs, and expanding envelopes of novae and supernovae. Synthetic spectra can be calculated in 1D and 3D using local thermal equilibrium (LTE) or non-LTE 
radiative transfer for any desired spectral resolution.

This new PHOENIX-ACES model grid was especially designed for modelling 
spectra of cool dwarfs, because it uses a new equation of state to improve the calculations of molecule formation in cool stellar atmospheres. This allows good fitting of the 
$\gamma$- and $\epsilon$-TiO bands ($\lambda_{\rm head}$ 705\,nm and $\lambda_{\rm head}$ 843\,nm, respectively), which are very sensitive to effective temperature. The $\epsilon$-TiO band is 
especially sensitive to temperatures lower than 3000\,K. The models use solar abundances from \cite{Asplund2009}. Models with [$\alpha$/Fe] $\neq 0$ are only available for $T_{\rm eff}$ > 3500\,K and 
$3 \leq$ [Fe/H] $\leq 0$. Therefore, we focus our analysis on models with [$\alpha$/Fe]=0. In this context, \cite{Veyette2016} reported a significant effect 
on the spectra of M dwarfs if abundances of other elements are varied. They found that a change in the C/O ratio influences the pseudo-continuum by changing TiO and H$_2$O opacities. 
In our study, however, we focused on the application of the latest PHOENIX-ACES models, with [Fe/H] being the only free abundance parameter.

\begin{figure}
  \includegraphics[width=0.45\textwidth]{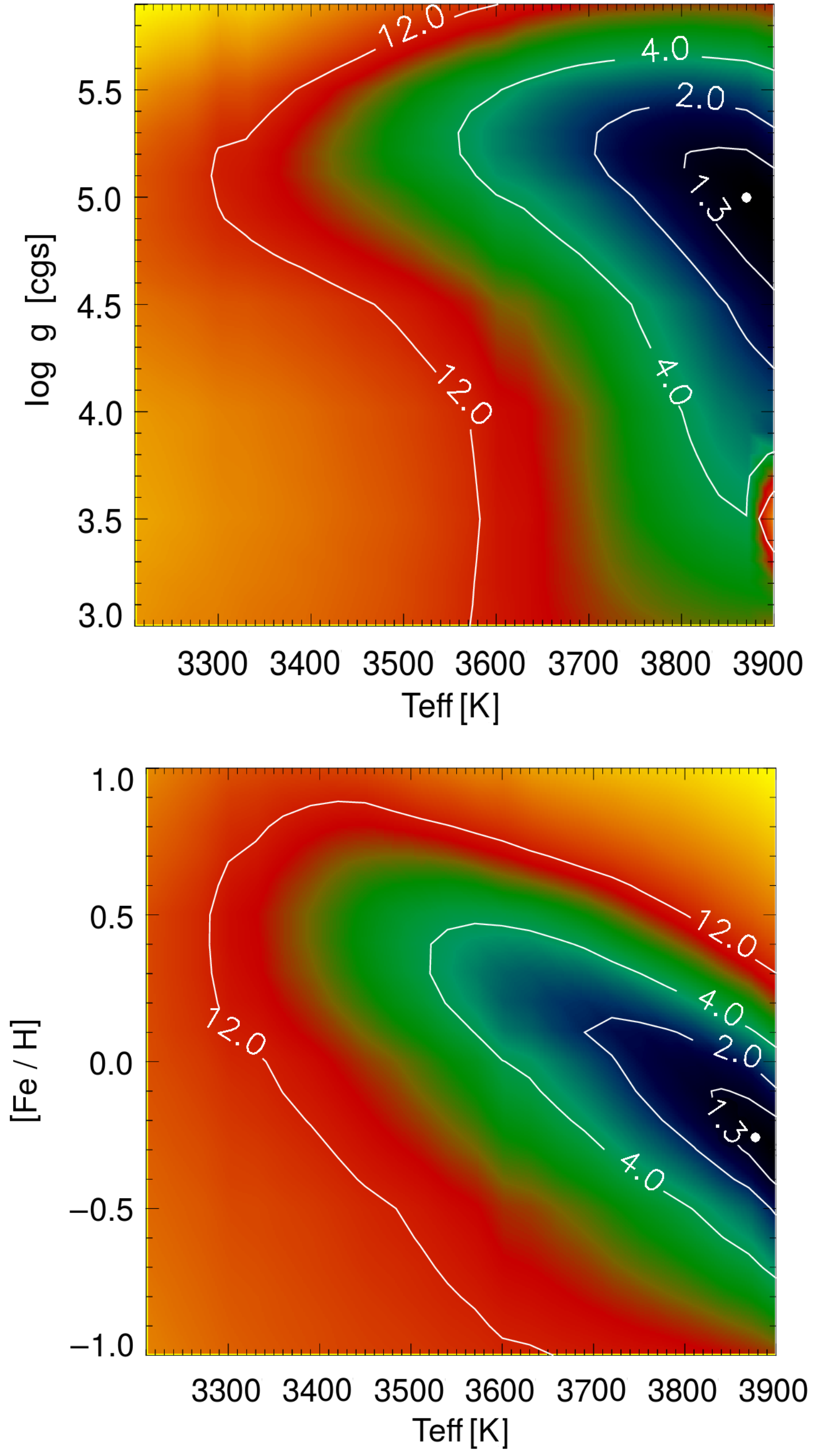}
  \caption{$\chi^2$ map of the Mg~{\sc i} line for \object{BR Psc} (GJ~908) with $\chi^2$ contour values. The upper panel shows the $T_{\rm eff}$-$\log{g}$ plane, and the lower panel 
  the $T_{\rm eff}$-[Fe/H] plane. The minimum for this line is indicated with a white dot.}
  \label{fig:Mg-chi2}
  \end{figure}

We slightly modified the algorithm developed by \cite{Passegger2016}. Because all stars in our sample have effective temperatures hotter than 3000\,K, we only included the $\gamma$-TiO band 
in our fitting. \cite{Passegger2016} also showed that the K~{\sc i} and Na~{\sc i} doublets around 768\,nm and 819\,nm, respectively, are suitable for surface gravity and metallicity 
determination. Since the K~{\sc i} line at 766.5\,nm is contaminated by telluric lines, we decided to exclude it from the fitting. We excluded the Ca~{\sc ii} doublet 
at 850.0\,nm and 866.4\,nm as well, because these lines are not well reproduced by the models: they 
are formed in the chromosphere and can show emission when the star is magnetically active.
The Na~{\sc i} doublet around 819\,nm was previously used because of its high pressure sensitivity \citep[see][]{Passegger2016}. In a detailed analysis of the first results of our 
large sample, we found a degeneracy in the strength and width of the Na~{\sc i} doublet over a wide parameter range, which made it difficult to distinguish 
between a cool-metal poor and a hot-metal rich model. Therefore we excluded this doublet from our analysis. The $\gamma$-TiO band and Mg~{\sc i} line ($\lambda$ 880.9\,nm) were found to be more 
suitable for metallicity determination, and they were therefore assigned higher weights during fitting. As an example, Fig.~\ref{fig:Mg-chi2} presents $\chi^2$ maps of the Mg~{\sc i} line for one of our 
stars in the $T_{\rm eff}$-$\log{g}$ and $T_{\rm eff}$-[Fe/H] plane, where a strong dependency on metallicity can be seen. 
A $\chi^2$ minimisation is used to determine the best fit of the models to the observed spectra. As described in \cite{Passegger2016}, the procedure is divided into two steps, which are described in 
the following. 

\subsection{Coarse grid search}
In a first step, we used the coarsely spaced grid of the model spectra in a wide range around the expected parameters of the star. To match the instrumental resolutions, the model spectra were first 
convolved with a Gaussian. Then the average flux of the models and the observed spectrum was normalised to unity by assuming a pseudo-continuum for each wavelength range. Next, the models 
were interpolated to match 
the wavelength grid of the observed spectrum, so that each wavelength point of each model spectrum could be compared to the stellar spectrum. The value of $\chi^2$ was calculated to find a rough 
global minimum. This was done for different wavelength ranges between 705.0 and 820.5\,nm. The parameters for the three best minima were given as an output in order to provide different 
starting values for the downhill simplex in the next step. 

\subsection{Fine grid search}
In the second step, the region around each global minimum was explored on a finer grid. The wavelength range was extended to 883.5\,nm  to include some titanium and iron lines. An overview of all 
regions and lines used for fitting is presented in Table~\ref{tab:lines}.
To reduce the number of free parameters in the fit, we used the values of projected 
rotation velocity $v\sin{i}$ determined by \cite{Jeffers2018} using cross-correlation. To account for $v\sin{i}$, the model spectra were broadened using a broadening function. 
The function determined the effect on the line spread function caused by stellar rotation. The resulting line spread function was convolved with the model spectrum. 
In contrast to \cite{Passegger2016}, who used the IDL {\tt curvefit} function, we used a downhill simplex method for fitting, which we found to be more robust on large samples. The downhill 
simplex used linear interpolation between the model grid points to explore the parameter space in detail. A $\chi^2$ minimisation finds the best-fit model. This was done for all 
three minima found in the previous step. The parameters with the best $\chi^2$ were selected as results.

%
%

\begin{figure*}
 \includegraphics[width=0.65\textwidth,bb = 30 10 840 430]{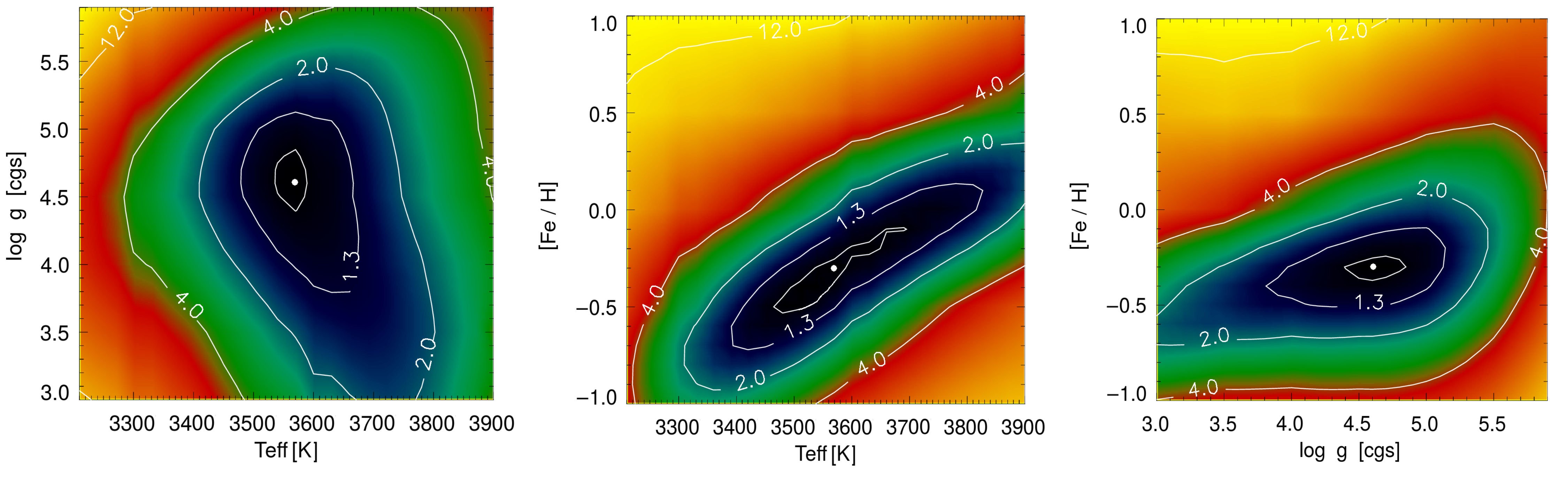}
 \caption{$\chi^2$ maps for \object{BR Psc} (GJ~908) for different combinations of stellar parameters. The global minimum is indicated with a white dot. }
 \label{fig:chi2-map}
 \end{figure*}

From the first results for our sample, the fits showed very good agreement between models and observed spectra. However, we found that the values of $\log{g}$ 
and [Fe/H] were much higher than expected for main-sequence M dwarfs; the $\log{g}$ was between 5.5 and 6.0, and most metallicities were super-solar, 
with values of up to 1.0\,dex. Moreover, we found exceptionally low $\log{g}$ of 3.0 with metallicities of $-1.0$\,dex for some stars. In both cases the 
fitted models agreed very well with the data. 
The results of obviously wrong parameter values can be explained by a degeneracy between $T_{\rm eff}$, $\log{g}$, and [Fe/H], which is displayed in Fig.~\ref{fig:chi2-map}. 
Especially the $T_{\rm eff}$-[Fe/H] map shows a largely extended minimum. To break this degeneracy, we decided to determine $\log{g}$ using an independent method.

\begin{figure*}
 \includegraphics[width=0.90\textwidth]{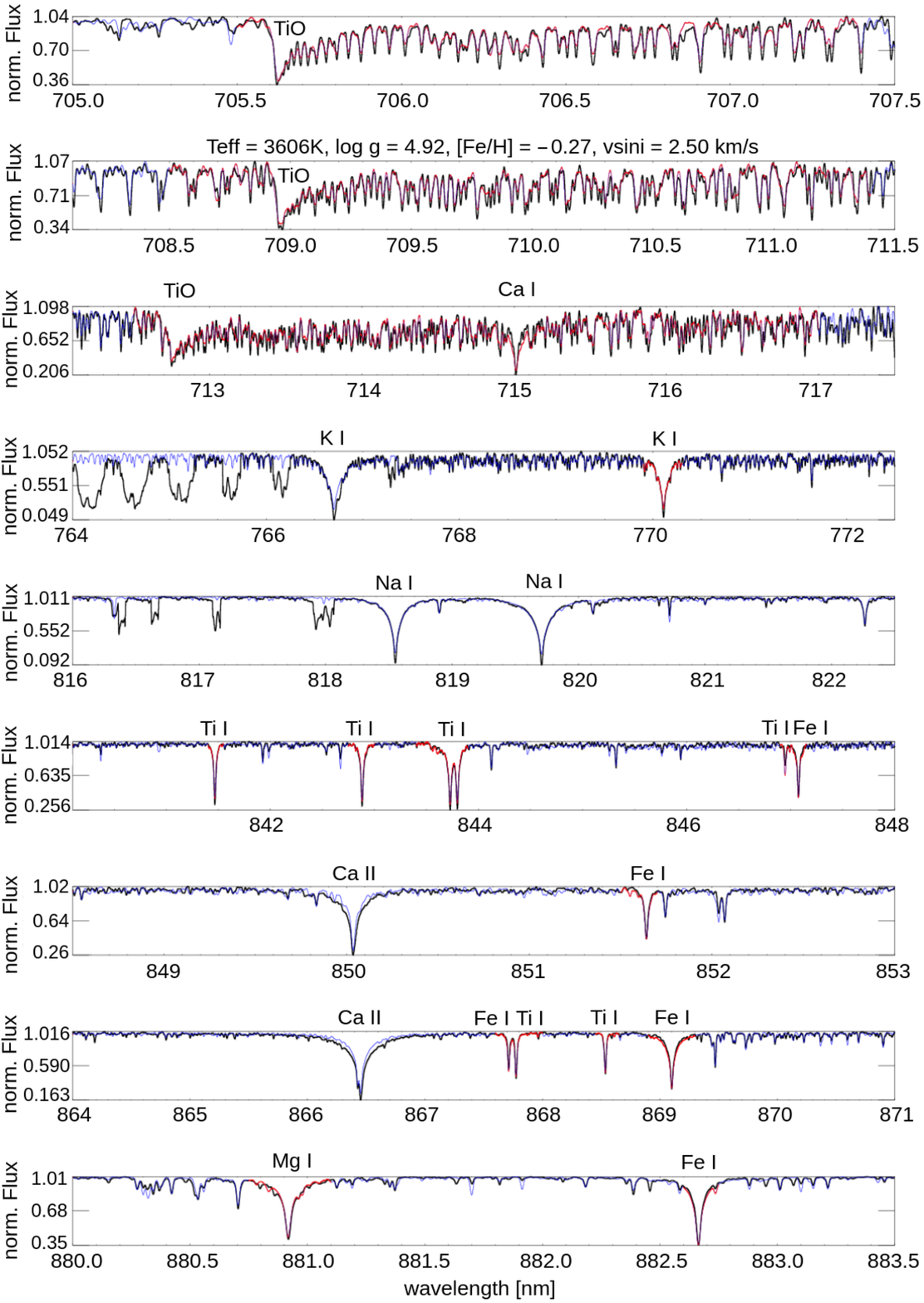}
 \caption{Co-added CARMENES spectrum of the M1\,V star \object{GX And} (black) and the best-fit model (blue: whole fit, and red: regions used for $\chi^2$ minimisation).}
 \label{fig:result}
 \end{figure*}

 \cite{Baraffe1998} presented evolutionary models for low-mass stars up to 1.4 M$_{\odot}$. A new version of these models was published by \cite{Baraffe2015} using updated solar 
abundances. However, the \cite{Baraffe1998} and \cite{Baraffe2015} $T_{\rm eff}$-$\log{g}$ relations are consistent with each other in the temperature range of M dwarfs, 
therefore we used the \cite{Baraffe1998} version. Amongst other parameters, they provided effective temperatures and surface gravities for different stellar ages and metallicities of 0.0 and $-$0.5\,dex. 
 Unfortunately, the ages of the stars in the CARMENES target sample are not yet well constrained. This will be the topic of upcoming papers. 
 A preliminary kinematics and activity analysis of the sample to qualitatively estimate ages was carried out by \cite{Cortes2016_2}. Therefore, we assumed an age of 5 Gyr for the whole sample. This 
 seems to be a good guess even for younger stars 
 because once M dwarfs reach the main sequence they evolve extremely slowly \citep[e.g.][]{Burrows1997,Laughlin1997}. This is also reflected in the \cite{Baraffe1998} relations, which agree within 
 0.02\,dex in $\log{g}$ for ages between 1 and 7 Gyr in the temperature range of M dwarfs up to 4000\,K. In the algorithm the downhill 
 simplex can vary $T_{\rm eff}$ and metallicity. Based on this, $\log{g}$ was determined from the $T_{\rm eff}$-$\log{g}$ relations. Metallicities between 0.0 and $-$0.5\,dex were linearly 
 interpolated from the relations to estimate $\log{g}$. For metallicities higher than 0.0\,dex or lower than $-$0.5\,dex, the values were extrapolated. Because the differences in $\log{g}$ 
 depending on metallicity are small (no larger than 0.20\,dex between metallicities 0.0 and $-$0.5\,dex), 
 we expect the uncertainty from the interpolation and extrapolation to be negligible compared to the 
 uncertainty coming from the fitting. From these three parameters, the corresponding PHOENIX model was interpolated and the $\chi^2$ was calculated.
Fig.~\ref{fig:result} shows a co-added CARMENES spectrum of a typical M1\,V star with the best-fit model, including the lines and regions we used for fitting.

\section{Results and discussion}

Table~A.1 presents the fundamental parameters of our target sample. It includes the CARMENES identifiers, spectral types from Carmencita, $T_{\rm eff}$, $\log{g}$, and [Fe/H] derived 
in this study, $v\sin{i}$ determined by \cite{Jeffers2018}, masses from Carmencita (see section 4.4), a flag for Ca~{\sc ii} emission, and the instrument with which the analysed spectrum was observed. 
We applied the method for error estimation as given in \cite{Passegger2016}. 
They estimated errors by adding Poisson noise to 1400 model spectra with random parameter distributions to simulate S/N $\sim$ 100 and applied their algorithm to recover the input 
parameters. Using this method, we derived uncertainties of 51\,K for $T_{\rm eff}$, 0.07\,dex for $\log{g}$, and 0.16\,dex for [Fe/H], which are consistent 
with typical uncertainties in literature. We confirmed this statement by calculating deviations between our results and literature values ($\sigma_{\rm exp}$) together with the corresponding 
standard deviation ($\sigma_{\Delta}$). The numbers are presented in Table~\ref{tab:errors}, showing that $\sigma_{\Delta}$ are smaller than the expected deviations $\sigma_{\rm exp}$ for the different 
literature samples. 

\begin{center}
  \begin{table}[H]
    \caption{Expected errors and deviations in $T_{\rm eff}$, $\log{g}$, and [Fe/H] of our results and the literature.}
    \label{tab:errors}
    \centering %
    \begin{tabular}{lcccccc}
      \hline \hline 
       Author$^{a}$ & \multicolumn{2}{c}{$T_{\rm eff}$ [K]} & \multicolumn{2}{c}{$\log{g}$ [dex]}  & \multicolumn{2}{c}{[Fe/H] [dex]} \\ 
       & $\sigma_{\rm{exp}}$ & $\sigma_{\Delta}$ & $\sigma_{\rm{exp}}$ & $\sigma_{\Delta}$ & $\sigma_{\rm{exp}}$ & $\sigma_{\Delta}$ \\
	   \hline 
      RA12 & 63 & 108 & ... & ... & 0.23 & 0.19 \\
      GM14 & 93 & 78 & ... & ... & 0.18 & 0.13 \\
      Ma15 & 85 & 51 & 0.09 & 0.08 & 0.18 & 0.10 \\
      \hline
      \multicolumn{7}{l}{$^{a}$ RA12: \cite{RojasAyala2012}, GM14: \cite{GaidosMann2014},} \\
      \multicolumn{7}{l}{Ma15: \cite{Maldonado2015}.}
    \end{tabular}
  \end{table}
\end{center}

\subsection{Effective temperature}
The histogram distributions for all parameters for all 300 stars are presented in Fig.~\ref{fig:histograms}. 
The temperature distribution (left panel) shows that most of the stars in our sample have temperatures of between 3200\,K and 3800\,K, corresponding to spectral types ranging from M0.0\,V to M5.0\,V. 
Figure~\ref{fig:comparison} gives a comparison of 98 stars that overlap with the samples of RA12, \citet[hereafter Ma15]{Maldonado2015}, and 
GM14. Ma15 determined effective temperature and metallicity from optical spectra using pseudo-equivalent widths. In general, most of our results 
agree with the literature values within the error bars. However, there is one group of outliers at the cool end of the sample. This group is represented by results from RA12, who determined temperatures using 
the H$_2$O-K2 index calibrated with BT-Settl models of solar metallicity. They derived temperatures that are cooler than ours by about 200\,K. Two more outliers are located around 3550\,K (GJ~752A) and 
3650\,K (BR~Psc/GJ~908), for which RA12 determined considerably hotter temperatures of 3789 and 3995\,K, respectively. However, our temperatures are consistent with those derived by Ma15 and GM14, 
which makes the result of RA12 discrepant. 
A small ``bump'' can be found between 3550 and 3700\,K, where GM14 tended to derive slightly higher temperatures than we do. 
For other stars in Fig.~\ref{fig:comparison}, the values are mostly consistent with our results.

\subsection{Surface gravity}
The middle panel of Fig.~\ref{fig:histograms} presents the $\log{g}$ distribution for our sample. Ma15 determined $\log{g}$ for early-M dwarfs using stellar masses from photometric relations 
and radii from an empirical mass-radius relation that combines interferometry \citep{vonBraun2014,Boyajian2012} and data from low-mass eclipsing binaries \citep{Hartman2015}. A comparison of 
those stars that we have in common is presented in the middle panel of Fig.~\ref{fig:comparison}. 
The grey lines indicate a 1\,$\sigma$ deviation of 0.07\,dex. We calculated $\log{g}$ for the sample of GM14 from the provided masses and radii. 
The uncertainties were derived with error propagation from the uncertainties in mass and radius. We included $\log{g}$ values based on interferometric radius measurements from \cite{Boyajian2012}.
We derived the $\log{g}$ in the same way as for GM14. 
Our results are consistent with the $\log{g}$ values from Ma15, which mostly lie between 4.6 and 5.0\,dex. Most of the interferometrically based $\log{g}$ also agree with our values. 
This is expected since they are constrained by the $T_{\rm eff}$-$\log{g}$ relations. It also shows consistency between the empirical radius calibration of Ma15 and theoretical models. 
However, when we compare our $\log{g}$ values with those of GM14, we find some offset. At the 
lower end of the plot, we derive higher values than GM14. Because $\log{g}$ depends on $T_{\rm eff}$ in our calculation, this trend is consistent with the bump found in the temperature plot. 
The values of \cite{Boyajian2012} slightly follow the same trend as GM14, although the sample is too small to draw a definite conclusion.

\subsection{Metallicity}
\begin{figure*}
 \includegraphics[width=0.7\textwidth,bb = 30 10 840 430]{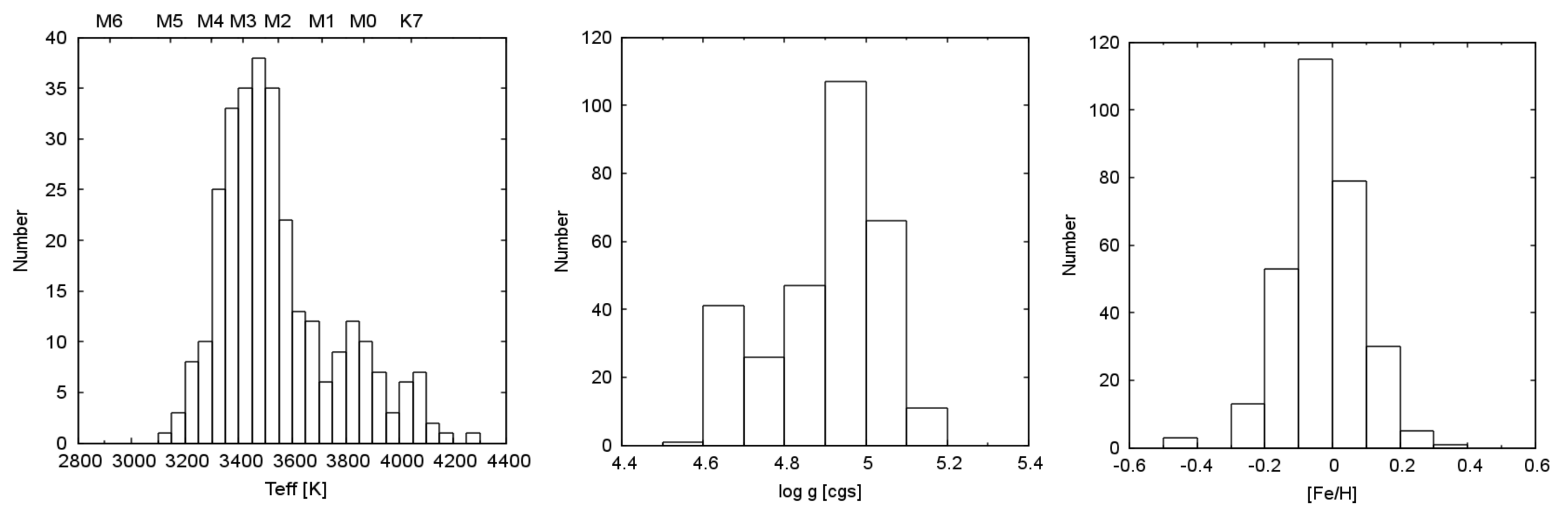}
 \caption{Histogram distributions of $T_{\rm eff}$ (left panel) together with spectral types from \cite{KenyonHartmann1995} on the upper x-axis, $\log{g}$ (middle panel), and 
 [Fe/H] (right panel) for our 300 stars.}
 \label{fig:histograms}
 \end{figure*}

\begin{figure*}[h]
 \includegraphics[width=0.7\textwidth,bb = 30 10 840 450]{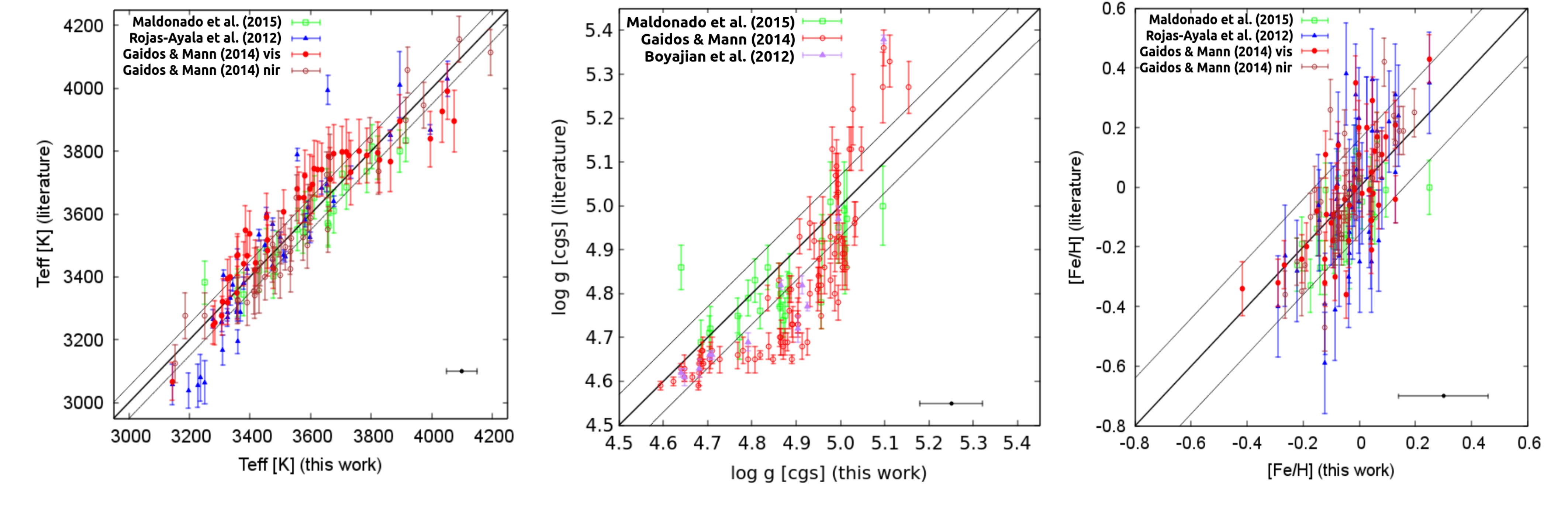}
 \caption{Comparison between values from our sample and literature values for $T_{\rm eff}$ (left panel), $\log{g}$ (middle panel), and [Fe/H] (right panel). The black line indicates the 1:1 relation. 
 The grey lines indicate the 1$\sigma$ deviation of 51\,K, 0.07\,dex in $\log{g}$, and 0.16\,dex in [Fe/H]. The black dots with error bars in the lower right corner of 
 each plot show the uncertainties of this work.}
 \label{fig:comparison}
 \end{figure*}

The right panel of Fig.~\ref{fig:histograms} displays the metallicity distribution of our results, centred on solar metallicity. 
The right panel of Fig.~\ref{fig:comparison} shows a comparison of the stars that we have in common with RA12, Ma15, 
and GM14. Their metallicity measurements range from --0.6 to 0.4\,dex, whereas our results only range from --0.4 to almost 0.2\,dex.
This indicates that the metallicity is more difficult to constrain than the other parameters, and that different methods could give noticeably different results. 
On the other hand, we find that even for spectra for which the parameters agree with the literature, some lines, such as Ti~{\sc i} ($\lambda$ 846.9\,nm and 867.77\,nm) and Fe~{\sc i} 
($\lambda$ 867.71\,nm and 882.6\,nm), are too deep. A possible explanation might be the contrast between the line and the continuum so that the models still cannot reproduce the 
correct line depths. On the other hand, we used models with element abundances fixed to solar. A change in [$\alpha$/Fe] or [Ti/Fe] might improve the fit for some stars. 
A more extensive study on the performance of the models themselves, probably including different element abundances, is necessary to completely understand their behaviour.

\subsection{Relation of spectral type, mass, and temperature}
\begin{figure}
 \includegraphics[width=0.5\textwidth]{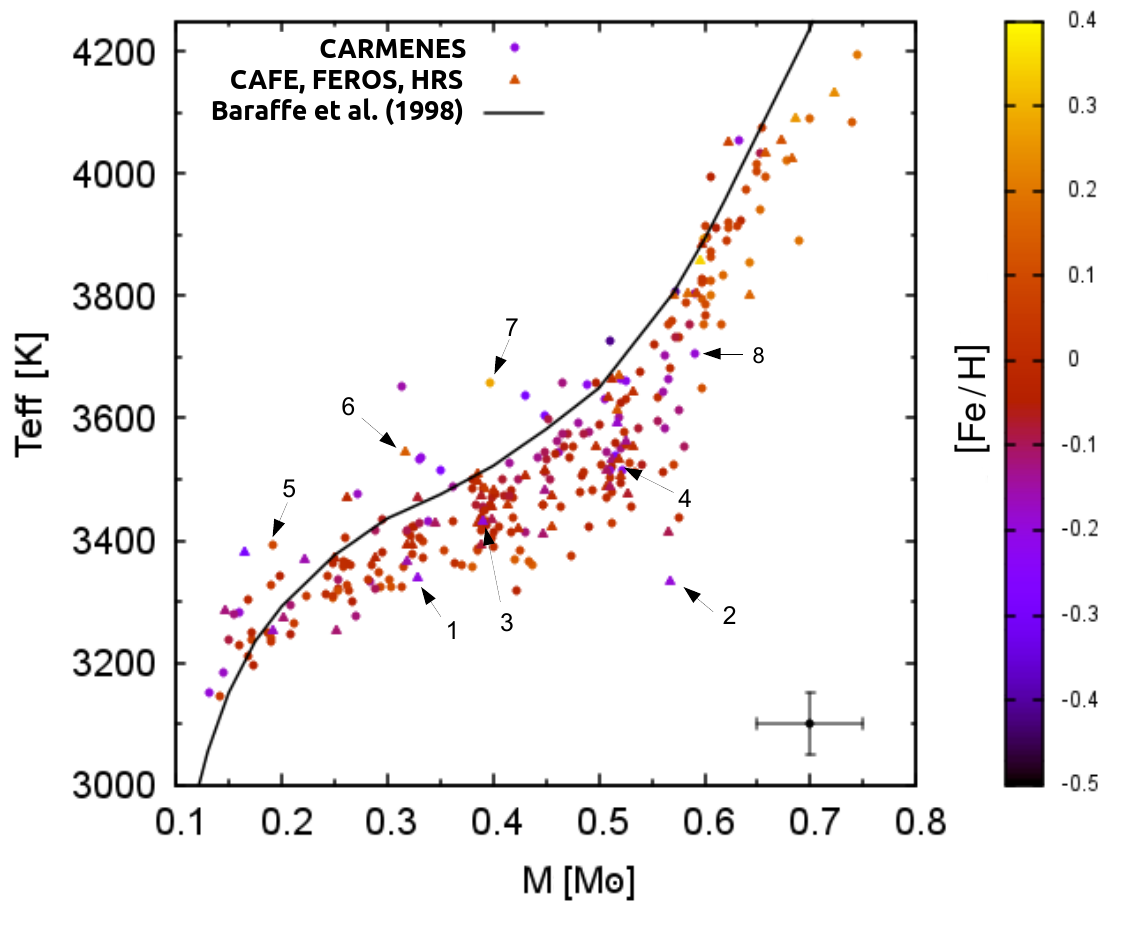}
 \caption{Effective temperature as a function of stellar mass. The determined metallicities of the stars are colour-coded. The different symbols present stars observed with CARMENES or 
 CAFE, FEROS, and HRS. The thick black line shows the theoretical relation for solar metallicity from \cite{Baraffe1998}. An average uncertainty of 10\%\ in mass is indicated by the 
 black dot with error bars in the lower right corner. The eight outliers discussed in Section 4.5 are labelled. }
 \label{fig:teff_mass}
 \end{figure}

We present the relation between stellar mass and effective temperature in Fig.~\ref{fig:teff_mass}, and the metallicities are colour-coded. The thick black line represents the theoretical 
relation from \cite{Baraffe1998} for an age of 5 Gyr and solar metallicity. The masses 
were calculated by combining mass-luminosity relations from \citet[for 4.5 mag < \textit{Ks} < 5.29 mag]{Delfosse2000} and \citet[for 5.29 mag < \textit{Ks} < 10 mag]{Benedict2016}, 
with the magnitudes taken from the Carmencita database \citep[see][]{Alonso2015}.
In this plot, stars with super-solar metallicity should lie below the relation reported by \cite{Baraffe1998} and stars with sub-solar metallicity should lie above this relation. 
As can be seen, most of the stars lie below the theoretical prediction. 
This can be due to several reasons: our $T_{\rm eff}$s are systematically underestimated, our metallicities are slightly lower than 
expected, or the determined stellar masses are overestimated. Based on the literature comparison in Section 4.1, we can exclude the former two. Since \cite{Delfosse2000} did not 
provide errors for their mass-luminosity relation, we assumed an average uncertainty of 10\% in mass over the whole mass range, which is of the same order as the errors from 
\cite{Benedict2016}. Within this range, our values agree with the theoretical relation of \cite{Baraffe1998}. Some obvious outliers are identified by numbers and are discussed in 
more detail later. 

\begin{figure}
 \includegraphics[width=0.5\textwidth]{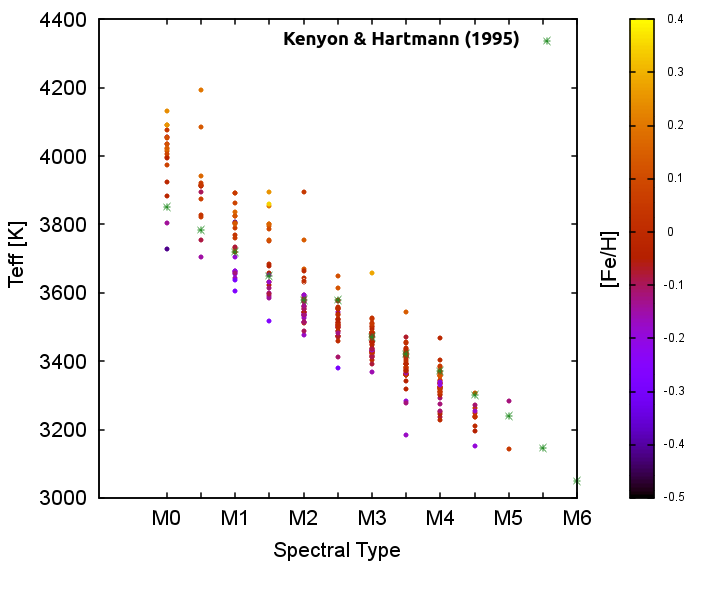}
 \caption{Effective temperature as a function of spectral type. The determined metallicities of the stars are colour-coded. The green stars indicate the expected 
 temperatures for each spectral type computed by \cite{KenyonHartmann1995}.}
 \label{fig:SpT-Teff}
 \end{figure}

Figure~\ref{fig:SpT-Teff} shows the effective temperatures of all stars as a function of their spectral type; the spectral types are taken from the Carmencita database. 
The green stars show the expected temperature-spectral type relation as presented by \cite{KenyonHartmann1995}. The authors computed effective temperatures, colours, spectral types, 
and bolometric corrections for main-sequence stars from B0 to M6 after an extensive literature search. Their temperatures fit our results for solar metallicities well. 
The large spread in temperature for each spectral type is caused by different metallicities, which are colour-coded. This indicates that stars of the 
same spectral type have higher temperatures if they are more metal-rich, or in other words: for the same effective temperature, the spectral type decreases with increasing 
metallicity. This can be 
explained with an increase in opacity in the optical with increasing metallicity, mainly dominated by TiO and VO molecular bands. The peak of the energy distribution is therefore 
shifted towards longer wavelengths and makes the star appear redder, that is, of later spectral type. This effect has been discussed in more detail by 
\cite{Delfosse2000,ChabrierBaraffe2000}. A similar trend was found by \cite{Mann2015}, who derived empirical relations between $T_{\rm eff}$, [Fe/H], radii, and luminosities. 
They showed that the radius increases with metallicity for a fixed temperature (see their Figure 23).

\subsection{Analysis of outliers}

In the following, eight outliers found in the mass-temperature plot of Fig.~\ref{fig:teff_mass} are discussed in more detail. We selected them because their $T_{\rm eff}$ 
or metallicity clearly deviate from the relation reported by \cite{Baraffe1998}. 

\paragraph{\textit{1: J03430+459.}} This star was observed with HRS. The best-fit model agrees moderately well with the observed spectrum, with small deviations in the TiO bandheads 
and some Ti~{\sc i} and Fe~{\sc i} lines. \cite{Alonso2015} measured a pseudo-equivalent width of --0.7 \AA{} for H$\alpha$. \cite{Newton2017} also reported that this star is slightly active. This 
could explain the deviations in the Fe~{\sc i} and Ti~{\sc i} lines, which are sensitive to magnetic Zeeman splitting. A change in [Ti/Fe] or [$\alpha$/Fe], on the other hand, could 
also be responsible for a deviation in the fit of Ti~{\sc i} lines.

\paragraph{\textit{2: J04544+650.}} For this star we also used HRS spectra to determine the parameters. The star is magnetically active, has an H$\alpha$ pseudo-equivalent width of --13.9 \AA{} 
\citep{Alonso2015}, and shows Ca~{\sc ii} emission. We find deviations in some Ti~{\sc i} and Fe~{\sc i} lines, which could explain the deviation in metallicity. 

\paragraph{\textit{3: J05078+179.}} This star was observed with FEROS. The best spectrum has an S/N of 104. The star shows H$\alpha$ emission with a pseudo-equivalent width of 
--0.7 \AA{} \citep{Jeffers2018}. The activity also causes distortion in other lines (e.g. some Fe~{\sc i} and Ti~{\sc i} lines). Here, the deviations in some Ti~{\sc i} lines might also be 
caused by Ti abundances that are different from solar, and this might in turn explain the low metallicity. 

\paragraph{\textit{4: J11201-104.}} We analysed CARMENES spectra of this star and found it to be too metal-poor in the mass-$T_{\rm eff}$ plot. The star has strong Ca~{\sc ii} emission. 
\cite{Alonso2015} reported an H$\alpha$ pseudo-equivalent width of --3.3 \AA{}. The magnetic activity might explain the deviations in the Fe~{\sc i} lines here as well, and therefore 
also the deviation in the determined lower metallicity. 

\paragraph{\textit{5: J18346+401.}} For this star we used a co-added CARMENES spectrum to derive the parameters. The best model fits the observed spectrum very well. The resulting temperature of 
3391\,K is comparable to the temperature measured by \cite{GaidosMann2014} from near-infrared data. However, \cite{GaidosMann2014} determined a metallicity of 0.42\,dex, whereas we 
obtained a more metal-poor value of 
0.09\,dex. The line depth is well fitted in our spectrum. We also derived a lower $\chi^2$ compared to the parameter set of \cite{GaidosMann2014}. The star is not known to be active and does 
not show any Ca~{\sc ii} emission either. Considering all this information, we cannot explain the measured low metallicity satisfactorily. 

\paragraph{\textit{6: J21057+502.}} We analysed HRS spectra for this star and found good agreement between the observed spectra and the best-fit models. Our derived temperature of 
3543\,K is about 100\,K too hot for the spectral type M3.5. However, the $\chi^2$-map shows a large extended minimum in the $T_{\rm eff}$-$\log{g}$-[Fe/H] planes, which reaches from 
almost 3700\,K and +0.5\,[Fe/H] down to 3500\,K and +0.1\,[Fe/H], making the derived parameters less significant. \cite{Cortes2016_2} reported from an analysis of the 
stellar kinematics that this star is part of the young disc and a probable member of the local association, which is 10-150 Myr old. The models of \cite{Baraffe1998} show that using an 
age of 5 Gyr for a star younger than 0.5 Gyr can lead to an increase in $T_{\rm eff}$ of 50--100\,K. Accounting for these two circumstances results in a slightly lower temperature and 
metallicity, which causes the star to fit the mass-$T_{\rm eff}$ relation better. 

\paragraph{\textit{7: J21152+257.}} The fit to this co-added CARMENES spectrum is good: we find only minor deviations between observed and fitted lines, especially for the TiO-band. 
The star is inactive and does not show any signs of emission in H$\alpha$ or Ca~{\sc {ii}}. The determined temperature of 3657\,K is about 200\,K hotter than expected for the stated 
spectral type M3. The $\chi^2$-map has a large extended and deep minimum here as well, which is located between about 3800\,K and +0.5\,[Fe/H], and 3550\,K and +0.1\,[Fe/H]. This 
might explain the too high temperature and metallicity. 

\paragraph{\textit{8: J21221+229.}} The parameters of this star were also derived from a co-added CARMENES spectrum, for which the model fit is very good. The star is not active, and the 
spectral type M1 corresponds to the fitted temperature of 3704\,K. We are unable to explain the deviation in Fig.~\ref{fig:teff_mass} for this star. 


\section{Summary}

CARMENES is a new instrument at the Calar Alto observatory that simultaneously takes high-resolution spectra in the visible and near-infrared wavelength ranges. 
Its aim is to search for Earth-sized planets in the habitable zone around M dwarfs.

We provided precise parameters from PHOENIX-ACES model fits for effective temperature, surface gravity, and metallicity for 300 M dwarfs, which is the largest sample 
of M dwarfs investigated with high-resolution spectroscopy so far.
It is important not only for CARMENES, but also for future exoplanet surveys, since knowing stellar fundamental parameters is essential for characterising an orbiting planet. 
Moreover, accurate metallicities are crucial for theories of planet formation around low-mass stars and give information on the chemical evolution of the Galaxy.

Our work presents a test of the new PHOENIX-ACES models on a large sample of low-mass stars and points out inconsistencies in line depths and metallicity determination.  
This analysis also serves as a comparison of methods using low- and high-resolution spectra for stellar parameter determination. 
Table~\ref{tab:lit} summarises the different literature approaches for determining stellar parameters. It illustrates that in contrast to other comparable studies, we used high-resolution 
spectra and fitting of the latest model atmospheres. 
Comparisons with literature values for some of the target stars showed that we achieve very good agreement in the temperatures. For the metallicity we find an overall distribution that shows 
mainly sub-solar values and peaks between 0.0 and $-$0.1\,dex, which agrees with findings from \citet[see their Figure 1]{GaidosMann2014}. Our values are consistent with the literature within 
1\,$\sigma$, although there is no obvious correlation between our values and literature results. This might indicate an inconsistency in metallicity determination as such and may require further 
improvement of methods and models. 
Simultaneous fitting of all three parameters did not provide reliable results for all sample stars. Therefore, we determined $\log{g}$ from temperature- and metallicity-dependent relations 
from evolutionary models assuming an average age of 5 Gyr for our sample. However, we showed that our results in $\log{g}$ agree well with interferometric 
observations by \cite{Boyajian2012}, which also serves as an evaluation of theoretical evolutionary models and observations. 

To confirm our results, we performed $\chi^2$ fits with our spectra and models with parameters determined by GM14, Ma15, and RA12. We compared the $\chi^2$s with those resulting from fits with our 
derived parameters and found the smallest $\chi^2$ for 92\%~of the fits with our parameters. For the remaining 8\%, the literature parameters agree with ours within their errors. This confirms  
that our method, using the latest PHOENIX-ACES models, provides the best-fit parameters to our observations. It shows that our method has the potential to derive accurate stellar parameters for 
M dwarfs. This contributes to the most extensive catalogue of M-dwarf parameters so far. 
However, we also showed that there are still some shortcomings in synthetic models for low-temperature atmospheres, although they have significantly improved in the past decade. 
While the PHOENIX-ACES models fit observed spectra very well and show only negligible deviations within the noise level, we can find some discrepancies. 
From the fit, the full line depth is not represented for some lines, which might be the reason for the differences in metallicity we found compared to literature values. 
A small offset in metallicity is also depicted in Fig.~\ref{fig:teff_mass}. Our results for solar metallicity lie systematically below the mass-luminosity relation of 
\cite{Baraffe1998}, but largely follow the theoretical prediction. Further detailed analysis of the models is necessary for better understanding the metallicity dependency. 

We identified eight outliers; four of them show activity either in H$\alpha$ or Ca~{\sc ii}. Magnetic activity can distort line profiles by Zeeman splitting 
\citep[e.g.][]{Hebrard2014,Reiners2013}, which could explain deviations in some sensitive Fe~{\sc i} and Ti~{\sc i} lines. Furthermore, a Ti and $\alpha$-element 
abundance different from the Sun might also cause deviations in the Ti~{\sc i} lines. Two outliers are caused by large extended and deep minima in the $T_{\rm eff}$-$\log{g}$-[Fe/H] planes. 
One of these stars is believed to have an age younger than 0.5\,Gyr \citep{Cortes2016_2}, which results in a too high temperature when using the models of \cite{Baraffe1998} for 
5\,Gyr. For two outliers we were not able to provide any explanation.

Finally, an accurate age determination of the sample stars would be helpful. This topic will be addressed in subsequent CARMENES papers. The S/N also has a significant influence 
on the parameter determination, which makes a high S/N preferable when using the method we presented here. 
A great advantage of the new CARMENES instrument is its capability to provide simultaneous observations in the visible and near-infrared wavelength range. A detailed investigation 
of spectra in both ranges is desirable to better understand M-dwarf atmospheres. The analysis of CARMENES near-infrared spectra will be presented in forthcoming works. 

\begin{center}
  \begin{table*}
    \caption{Summary of literature approaches for the stellar parameter determination.}
    \label{tab:lit}
    \centering %
    \begin{tabular}{lccccc}
      \hline \hline 
       Author$^{a}$ & Resolution & $\Delta \lambda$ [nm] & $T_{\rm eff}$ & $\log{g}$ & [Fe/H]\\
	   \hline 
      RA12 & \textasciitilde 2700  & 1000-2400 & H$_2$O-K2 index & ... & Na~{\sc i} and Ca~{\sc i} EW,\\
	& & & & & H$_2$O-K2 index \\
      GM14 & 800 to 1000 &  320-970 & BT-Settl fit, & ... & atomic line\\
       & 2000 & 800-2400 & spec. curvature & & strength relation \\
      Ma15 & 115,000 & 378-693 & pseudo-EW  & masses and radii & pseudo-EW \\
       & & & & from empirical relation & \\
      This work & 48000 to& 700-880 & PHOENIX-ACES fit& \cite{Baraffe1998} & PHOENIX-ACES fit \\
       &94600& & with downhill simplex& relation & with downhill simplex\\
      \hline
      \multicolumn{6}{l}{$^{a}$ RA12: \cite{RojasAyala2012}, GM14: \cite{GaidosMann2014}, Ma15: \cite{Maldonado2015}.}
    \end{tabular}
  \end{table*}
\end{center}


\begin{acknowledgements}

We thank the anonymous referee for her/his comments that helped to improve the quality of this paper.
CARMENES is an instrument for the Centro Astron\'omico Hispano-Alem\'an de Calar Alto (CAHA, Almer\'{\i}a, Spain). 
CARMENES is funded by the German Max-Planck-Gesellschaft (MPG), the Spanish Consejo Superior de Investigaciones Cient\'{\i}ficas (CSIC), 
the European Union through FEDER/ERF FICTS-2011-02 funds, and the members of the CARMENES Consortium (Max-Planck-Institut f\"ur Astronomie, Instituto de Astrof\'{\i}sica de Andaluc\'{\i}a, 
Landessternwarte K\"onigstuhl, Institut de Ci\`encies de l'Espai, Institut f\"ur Astrophysik G\"ottingen, Universidad Complutense de Madrid, Th\"uringer Landessternwarte Tautenburg, 
Instituto de Astrof\'{\i}sica de Canarias, Hamburger Sternwarte, Centro de Astrobiolog\'{\i}a and Centro Astron\'omico Hispano-Alem\'an), with additional contributions by the Spanish 
Ministry of Economy, the German Science Foundation through the Major Research Instrumentation Programme and DFG Research Unit FOR2544 ``Blue Planets around Red Stars'', 
the Klaus Tschira Stiftung, the states of Baden-W\"urttemberg and Niedersachsen, and by the Junta de Andaluc\'{\i}a.
IR acknowledges support from the Spanish Ministry of Economy and Competitiveness (MINECO) through grant ESP2014-57495-C2-2-R.
VJSB is supported by programme AYA2015-69350-C3-2-P from Spanish Ministry of Economy and Competitiveness (MINECO)
Based on observations collected at the Centro Astron\'omico Hispano Alem\'an (CAHA) at Calar Alto, operated jointly by the Max--Planck Institut f\"ur Astronomie and the Instituto de 
Astrof\'{\i}sica de Andaluc\'{\i}a.
This research has made use of the VizieR catalogue access tool, CDS, Strasbourg, France. The original description of the VizieR service was published in A\&AS 143, 23

\end{acknowledgements}

\bibliographystyle{aa} 
\bibliography{hires}

\newpage

\appendix

\section{Table of parameters}
The online version contains full names and equatorial coordinates of all stars. The electronic form is available 
at the CDS via anonymous ftp to cdsarc.u-strasbg.fr (130.79.128.5) or via http://cdsweb.u-strasbg.fr/cgi-bin/qcat?J/A+A/. \newline
Columns and references to $v \sin i$ are discussed in a footnote below the table. 


\begin{longtable}{llcccccll}
\label{tab:results}\\
\caption[]{Basic astrophysical parameters of investigated stars$^{a}$.}\\
   \hline
   \hline
   \noalign{\smallskip}
Karmn 	&  Spectral	& $T_{\rm eff}$ [K]	& $\log{g}$ [dex]	& [Fe/H] [dex]	& $v \sin i$ & M & Ca~{\sc ii} IRT	& Instrument  \\ 
      		& type		& ($\pm$\,51\,K)  & ($\pm$\,0.07\,dex) & ($\pm$\,0.16\,dex) & [km/s]& [M$_\odot$] &	emission		&  	\\
\noalign{\smallskip}
    \hline
    \noalign{\smallskip}		
 \endfirsthead
\caption[]{Basic astrophysical parameters of investigated stars$^{a}$ (cont.).}\\ 
  \hline
  \hline
  \noalign{\smallskip}		
Karmn 	&  Spectral	& $T_{\rm eff}$ [K]	& $\log{g}$ [dex]	& [Fe/H] [dex]	& $v \sin i$ & M  & Ca~{\sc ii} IRT	& Instrument  \\ 
      		& type		& ($\pm$\,51\,K)  & ($\pm$\,0.07\,dex) & ($\pm$\,0.16\,dex) & [km/s] & [M$_\odot$]&	emission		&  	\\
  \noalign{\smallskip}
  \hline
  \noalign{\smallskip}
  \endhead
  \noalign{\smallskip}
  \hline
  \endfoot
J00051+457	&	M1.0\,V	&	3665	&	4.85	&	--0.16	&	< 3	&	0.565	&	…	&	CARM	co-add	\\
J00056+458	&	M0.0\,V	&	4055	&	4.64	&	+0.11	&	< 3	&	0.672	&	…	&	CAFE	\\	
J00162+198E	&	M4.0\,V	&	3336	&	5.02	&	0.08	&	< 3	&	0.302	&	…	&	CARM	co-add	\\
J00183+440	&	M1.0\,V	&	3606	&	4.93	&	--0.27	&	2.5$^{b}$	&	0.449	&	…	&	CARM	co-add	\\
J00184+440	&	M3.5\,V	&	3283	&	5.11	&	--0.19	&	1.9$^{c}$	&	0.159	&	…	&	CARM	co-add	\\
J00286--066	&	M4.0\,V	&	3387	&	4.99	&	0.05	&	< 3	&	0.385	&	…	&	CARM	co-add	\\
J00315--058	&	M3.5\,V	&	3392	&	5.01	&	--0.02	&	< 3	&	0.323	&	…	&	FEROS	\\	
J00389+306	&	M2.5\,V	&	3537	&	4.89	&	--0.04	&	2.5$^{b}$	&	0.472	&	…	&	CARM	co-add	\\
J00395+149S	&	M4.0\,V	&	3334	&	5.06	&	--0.09	&	< 3	&	0.332	&	…	&	HRS	\\	
J00570+450	&	M3.0\,V	&	3425	&	4.99	&	--0.05	&	< 3	&	0.394	&	…	&	CARM	co-add	\\
J01013+613	&	M2.0\,V	&	3537	&	4.92	&	--0.13	&	4.0$^{d}$	&	0.442	&	…	&	CARM	co-add	\\
J01025+716	&	M3.0\,V	&	3478	&	4.92	&	0.00	&	2.5$^{b}$	&	0.512	&	…	&	CARM	co-add	\\
J01026+623	&	M1.5\,V	&	3796	&	4.69	&	0.13	&	< 3	&	0.597	&	yes	&	CARM	co-add	\\
J01125--169	&	M4.5\,V	&	3152	&	5.17	&	--0.20	&	2.5$^{b}$	&	0.132	&	…	&	CARM	co-add	\\
J01339--176	&	M4.0\,V	&	3335	&	5.07	&	--0.11	&	< 3	&	0.254	&	…	&	CARM	co-add	\\
J01384+006	&	M2.0\,V	&	3644	&	4.80	&	0.01	&	< 3	&	0.532	&	…	&	FEROS	\\	
J01433+043	&	M2.0\,V	&	3534	&	4.91	&	--0.08	&	2.5$^{b}$	&	0.451	&	…	&	CARM	co-add	\\
J01518+644	&	M2.5\,V	&	3553	&	4.89	&	--0.06	&	4.0$^{d}$	&	0.467	&	…	&	CARM	co-add	\\
J02002+130	&	M3.5\,V	&	3185	&	5.15	&	--0.18	&	< 3	&	0.144	&	…	&	CARM	co-add	\\
J02015+637	&	M3.0\,V	&	3495	&	4.93	&	--0.05	&	2.5$^{b}$	&	0.521	&	…	&	CARM	co-add	\\
J02026+105	&	M4.5\,V	&	3254	&	5.12	&	--0.17	&	6.00	&	0.191	&	yes	&	FEROS	\\	
J02050--176	&	M2.5\,V	&	3534	&	4.88	&	0.00	&	< 3	&	0.519	&	…	&	FEROS	\\	
J02070+496	&	M3.5\,V	&	3414	&	5.02	&	--0.12	&	< 3	&	0.431	&	…	&	CARM	co-add	\\
J02096--143	&	M2.5\,V	&	3555	&	4.87	&	0.00	&	< 3	&	0.533	&	…	&	FEROS	\\	
J02116+185	&	M3.0\,V	&	3428	&	4.97	&	0.00	&	< 3	&	0.385	&	…	&	FEROS	\\	
J02123+035	&	M1.5\,V	&	3659	&	4.81	&	--0.05	&	< 3	&	0.497	&	…	&	CARM	co-add	\\
J02222+478	&	M0.5\,V	&	3921	&	4.68	&	0.06	&	4	&	0.622	&	…	&	CARM	co-add	\\
J02336+249	&	M4.0\,V	&	3293	&	5.09	&	--0.11	&	3.1	&	0.208	&	yes	&	CARM	co-add	\\
J02358+202	&	M2.0\,V	&	3595	&	4.88	&	--0.10	&	< 3	&	0.555	&	…	&	CARM	co-add	\\
J02362+068	&	M4.0\,V	&	3326	&	5.03	&	0.04	&	< 3	&	0.261	&	…	&	CARM	co-add	\\
J02442+255	&	M3.0\,V	&	3459	&	4.96	&	--0.07	&	2.5$^{b}$	&	0.384	&	…	&	CARM	co-add	\\
J02565+554W	&	M1.0\,V	&	3891	&	4.66	&	0.19	&	4.0$^{d}$	&	0.689	&	…	&	CARM	co-add	\\
J02581--128	&	M2.5\,V	&	3381	&	5.08	&	--0.30	&	< 3	&	0.165	&	…	&	FEROS	\\	
J03026--181	&	M2.5\,V	&	3613	&	4.78	&	0.12	&	< 3	&	0.517	&	…	&	FEROS	\\	
J03181+382	&	M1.5\,V	&	3854	&	4.66	&	0.20	&	2.5$^{c}$	&	0.642	&	…	&	CARM	co-add	\\
J03213+799	&	M2.0\,V	&	3574	&	4.90	&	--0.11	&	4.0$^{d}$	&	0.465	&	…	&	CARM	co-add	\\
J03217--066	&	M2.0\,V	&	3552	&	4.91	&	--0.13	&	< 3	&	0.521	&	yes	&	CARM	co-add	\\
J03233+116	&	M2.5\,V	&	3412	&	5.02	&	--0.11	&	< 3	&	0.447	&	yes	&	FEROS	\\	
J03430+459	&	M4.0\,V	&	3338	&	5.08	&	--0.20	&	< 3	&	0.329	&	…	&	HRS	\\	
J03438+166	&	M0.0\,V	&	4034	&	4.64	&	0.12	&	< 3	&	0.657	&	…	&	FEROS	\\	
J03463+262	&	M0.0\,V	&	3997	&	4.65	&	0.11	&	< 3	&	0.658	&	yes	&	CARM	co-add	\\
J03531+625	&	M3.0\,V	&	3484	&	4.94	&	--0.04	&	< 3	&	0.380	&	…	&	CARM	co-add	\\
J04225+105	&	M3.5\,V	&	3438	&	4.96	&	0.00	&	< 3	&	0.575	&	…	&	CARM	co-add	\\
J04290+219	&	M0.5\,V	&	4194	&	4.59	&	0.20	&	1.11$^{e}$	&	0.744	&	…	&	CARM	co-add	\\
J04311+589	&	M4.0\,V	&	3325	&	5.03	&	0.05	&	< 3	&	0.313	&	…	&	CARM	co-add	\\
J04376--110	&	M1.5\,V	&	3624	&	4.84	&	--0.05	&	< 3	&	0.520	&	…	&	CARM	co-add	\\
J04376+528	&	M0.0\,V	&	4034	&	4.68	&	--0.09	&	< 3	&	0.653	&	yes	&	CARM	co-add	\\
J04429+189	&	M2.0\,V	&	3582	&	4.88	&	--0.08	&	< 3	&	0.537	&	…	&	CARM	co-add	\\
J04429+214	&	M3.5\,V	&	3424	&	4.98	&	0.00	&	< 3	&	0.323	&	…	&	CARM	co-add	\\
J04520+064	&	M3.5\,V	&	3391	&	5.00	&	0.00	&	2.5$^{b}$	&	0.400	&	…	&	CARM	co-add	\\
J04538--177	&	M2.0\,V	&	3563	&	4.90	&	--0.12	&	2.5$^{b}$	&	0.460	&	…	&	CARM	\\	
J04544+650	&	M4.0\,V	&	3332	&	5.09	&	--0.19	&	< 3	&	0.568	&	yes	&	HRS	\\	
J04588+498	&	M0.0\,V	&	4015	&	4.65	&	0.09	&	< 3	&	0.649	&	yes	&	CARM	co-add	\\
J05033--173	&	M3.0\,V	&	3416	&	5.01	&	--0.10	&	2.5$^{b}$	&	0.288	&	…	&	CARM	\\	
J05050+442	&	M5.0\,V	&	3285	&	5.10	&	--0.12	&	< 3	&	0.146	&	…	&	HRS	\\	
J05078+179	&	M3.0\,V	&	3432	&	5.02	&	--0.20	&	3	&	0.391	&	…	&	FEROS	\\	
J05091+154	&	M3.0\,V	&	3412	&	5.01	&	--0.09	&	4	&	0.565	&	yes	&	FEROS	\\	
J05127+196	&	M2.0\,V	&	3579	&	4.89	&	--0.12	&	2.5$^{b}$	&	0.491	&	…	&	CARM	co-add	\\
J05280+096	&	M3.5\,V	&	3362	&	5.03	&	--0.03	&	< 3	&	0.249	&	…	&	CARM	co-add	\\
J05298--034	&	M2.5\,V	&	3474	&	4.93	&	0.00	&	< 3	&	0.455	&	…	&	FEROS	\\	
J05314--036	&	M1.5\,V	&	3894	&	4.64	&	0.25	&	< 3$^{f}$	&	0.599	&	…	&	CARM	co-add	\\
J05348+138	&	M3.5\,V	&	3424	&	4.98	&	0.00	&	2.5$^{b}$	&	0.405	&	…	&	CARM	co-add	\\
J05360--076	&	M4.0\,V	&	3365	&	5.01	&	0.01	&	4.0$^{d}$	&	0.259	&	…	&	CARM	co-add	\\
J05365+113	&	M0.0\,V	&	4075	&	4.65	&	0.04	&	6.40	&	0.655	&	yes	&	CARM	co-add	\\
J05366+112	&	M4.0\,V	&	3333	&	5.07	&	--0.14	&	< 3	&	0.283	&	yes	&	CARM	\\	
J05415+534	&	M1.0\,V	&	3863	&	4.69	&	0.08	&	2.0$^{c}$	&	0.605	&	…	&	CARM	co-add	\\
J05421+124	&	M4.0\,V	&	3310	&	5.05	&	0.04	&	< 3	&	0.223	&	…	&	CARM	co-add	\\
J05532+242	&	M1.5\,V	&	3755	&	4.71	&	0.11	&	< 3	&	0.616	&	…	&	CARM	co-add	\\
J06011+595	&	M3.5\,V	&	3358	&	5.02	&	0.00	&	< 3	&	0.265	&	…	&	CARM	co-add	\\
J06103+821	&	M2.0\,V	&	3543	&	4.89	&	--0.05	&	2.5$^{b}$	&	0.458	&	…	&	CARM	co-add	\\
J06105--218	&	M0.5\,V	&	3822	&	4.71	&	0.06	&	1.0$^{f}$	&	0.598	&	…	&	CARM	co-add	\\
J06246+234	&	M4.0\,V	&	3238	&	5.11	&	--0.08	&	< 3	&	0.150	&	…	&	CARM	co-add	\\
J06277+093	&	M2.0\,V	&	3534	&	4.92	&	--0.12	&	< 3	&	0.513	&	…	&	FEROS	\\	
J06325+641	&	M4.0\,V	&	3469	&	4.94	&	--0.01	&	< 3	&	0.261	&	…	&	HRS	\\	
J06371+175	&	M0.0\,V	&	3728	&	4.89	&	--0.42	&	< 3	&	0.510	&	…	&	CARM	co-add	\\
J06396--210	&	M4.0\,V	&	3322	&	5.06	&	--0.04	&	3.70	&	0.253	&	…	&	CARM	\\	
J06421+035	&	M3.5\,V	&	3436	&	4.96	&	0.02	&	< 3	&	0.419	&	…	&	CARM	\\	
J06548+332	&	M3.0\,V	&	3450	&	4.96	&	--0.02	&	< 3	&	0.392	&	…	&	CARM	co-add	\\
J07033+346	&	M4.0\,V	&	3276	&	5.10	&	--0.12	&	3.50	&	0.270	&	yes	&	CARM	co-add	\\
J07044+682	&	M3.0\,V	&	3469	&	4.94	&	--0.01	&	< 3	&	0.418	&	…	&	CARM	co-add	\\
J07081--228	&	M2.0\,V	&	3664	&	4.79	&	--0.01	&	< 3	&	0.512	&	…	&	FEROS	\\	
J07274+052	&	M3.5\,V	&	3358	&	5.01	&	0.04	&	< 3	&	0.315	&	…	&	CARM	co-add	\\
J07287--032	&	M3.0\,V	&	3458	&	4.95	&	--0.02	&	2.5$^{b}$	&	0.447	&	…	&	CARM	co-add	\\
J07319+362N	&	M3.5\,V	&	3319	&	5.06	&	--0.03	&	< 3	&	0.422	&	yes	&	CARM	co-add	\\
J07349+147	&	M3.0\,V	&	3435	&	5.00	&	--0.09	&	4.8	&	0.398	&	yes	&	FEROS	\\	
J07353+548	&	M2.0\,V	&	3526	&	4.93	&	--0.14	&	< 3	&	0.415	&	…	&	CARM	co-add	\\
J07361--031	&	M1.0\,V	&	3891	&	4.69	&	0.05	&	3.5	&	0.621	&	yes	&	CARM	co-add	\\
J07386--212	&	M3.0\,V	&	3417	&	5.00	&	--0.09	&	< 3	&	0.319	&	…	&	CARM	co-add	\\
J07393+021	&	M0.0\,V	&	4005	&	4.66	&	0.07	&	< 3	&	0.650	&	yes	&	CARM	co-add	\\
J07545+085	&	M2.5\,V	&	3483	&	4.96	&	--0.13	&	< 3	&	0.448	&	…	&	FEROS	\\	
J07582+413	&	M3.5\,V	&	3363	&	5.02	&	0.00	&	< 3	&	0.262	&	…	&	CARM	co-add	\\
J08126--215	&	M4.0\,V	&	3326	&	5.03	&	0.04	&	< 3	&	0.189	&	…	&	CARM	\\	
J08161+013	&	M2.0\,V	&	3589	&	4.86	&	--0.06	&	< 3	&	0.500	&	…	&	CARM	co-add	\\
J08293+039	&	M2.5\,V	&	3575	&	4.88	&	--0.07	&	< 3	&	0.470	&	…	&	CARM	\\	
J08313--060	&	M1.5\,V	&	3802	&	4.68	&	0.16	&	< 3	&	0.642	&	…	&	FEROS	\\	
J08344--011	&	M3.5\,V	&	3371	&	5.02	&	--0.03	&	< 3	&	0.250	&	…	&	FEROS	\\	
J08358+680	&	M2.5\,V	&	3471	&	4.95	&	--0.06	&	< 3	&	0.399	&	…	&	CARM	\\	
J08371+151	&	M2.5\,V	&	3489	&	4.92	&	0.00	&	< 3	&	0.507	&	…	&	FEROS	\\	
J08402+314	&	M3.5\,V	&	3381	&	5.02	&	--0.04	&	< 3	&	0.295	&	…	&	CARM	\\	
J08427+095	&	M0.0\,V	&	4024	&	4.64	&	0.14	&	< 3	&	0.682	&	…	&	FEROS	\\	
J08428+095	&	M2.5\,V	&	3505	&	4.91	&	--0.01	&	< 3	&	0.430	&	…	&	FEROS	\\	
J08526+283	&	M4.5\,V	&	3307	&	5.03	&	0.13	&	2.5$^{b}$	&	0.248	&	…	&	CARM	\\	
J08551+015	&	M0.0\,V	&	4091	&	4.61	&	0.25	&	< 3	&	0.686	&	…	&	FEROS	\\	
J09008+052E	&	M3.5\,V	&	3457	&	4.94	&	0.02	&	< 3	&	0.414	&	…	&	FEROS	\\	
J09008+052W	&	M3.0\,V	&	3424	&	4.97	&	0.05	&	< 3	&	0.455	&	…	&	FEROS	\\	
J09023+084	&	M2.5\,V	&	3507	&	4.90	&	0.00	&	< 3	&	0.521	&	…	&	FEROS	\\	
J09028+680	&	M4.0\,V	&	3343	&	5.03	&	0.01	&	4.0$^{d}$	&	0.244	&	…	&	CARM	\\	
J09133+688	&	M2.5\,V	&	3545	&	4.93	&	--0.16	&	< 3	&	0.462	&	yes	&	CARM	\\	
J09143+526	&	M0.0\,V	&	4053	&	4.65	&	0.07	&	< 3	&	0.622	&	…	&	CAFE	\\	
J09144+526	&	M0.0\,V	&	3994	&	4.68	&	--0.03	&	3.21$^{g}$	&	0.605	&	yes	&	CARM	co-add	\\
J09163--186	&	M1.5\,V	&	3584	&	4.90	&	--0.14	&	< 3	&	0.563	&	…	&	CARM	\\	
J09288--073	&	M2.5\,V	&	3496	&	4.91	&	0.00	&	< 3	&	0.385	&	…	&	FEROS	\\	
J09307+003	&	M3.5\,V	&	3413	&	4.99	&	--0.01	&	< 3	&	0.319	&	…	&	CARM	\\	
J09360--216	&	M2.5\,V	&	3488	&	4.96	&	--0.14	&	2.5$^{b}$	&	0.362	&	…	&	CARM	\\	
J09411+132	&	M1.5\,V	&	3601	&	4.88	&	--0.14	&	< 3	&	0.519	&	…	&	CARM	co-add	\\
J09423+559	&	M3.5\,V	&	3384	&	4.99	&	0.07	&	< 3	&	0.425	&	…	&	CARM	\\	
J09425+700	&	M2.0\,V	&	3511	&	4.91	&	--0.03	&	10.0$^{h}$	&	0.560	&	yes	&	CARM	co-add	\\
J09428+700	&	M3.0\,V	&	3423	&	4.99	&	--0.04	&	2.5$^{b}$	&	0.491	&	…	&	CARM	co-add	\\
J09468+760	&	M1.5\,V	&	3683	&	4.78	&	0.00	&	< 3	&	0.568	&	…	&	CARM	co-add	\\
J09511--123	&	M0.5\,V	&	3753	&	4.77	&	--0.09	&	< 3	&	0.585	&	…	&	CARM	co-add	\\
J09561+627	&	M0.0\,V	&	3974	&	4.67	&	0.07	&	< 3	&	0.640	&	yes	&	CARM	co-add	\\
J10023+480	&	M1.0\,V	&	3768	&	4.73	&	0.03	&	< 3	&	0.601	&	…	&	CARM	co-add	\\
J10087+027	&	M3.0\,V	&	3486	&	4.90	&	0.06	&	< 3	&	0.392	&	…	&	FEROS	\\	
J10122--037	&	M1.5\,V	&	3613	&	4.87	&	--0.12	&	< 3	&	0.575	&	…	&	CARM	co-add	\\
J10125+570	&	M3.5\,V	&	3408	&	4.99	&	--0.01	&	< 3	&	0.321	&	…	&	CARM	\\	
J10158+174	&	M3.5\,V	&	3392	&	5.01	&	--0.02	&	< 3	&	0.319	&	…	&	FEROS	\\	
J10167--119	&	M3.0\,V	&	3511	&	4.89	&	0.01	&	< 3	&	0.534	&	…	&	CARM	co-add	\\
J10243+119	&	M2.0\,V	&	3488	&	4.95	&	--0.10	&	< 3	&	0.511	&	…	&	FEROS	\\	
J10251--102	&	M1.0\,V	&	3761	&	4.73	&	0.05	&	< 3	&	0.569	&	…	&	CARM	co-add	\\
J10289+008	&	M2.0\,V	&	3575	&	4.89	&	--0.09	&	< 3	&	0.485	&	…	&	CARM	co-add	\\
J10350--094	&	M3.0\,V	&	3457	&	4.95	&	--0.03	&	< 3	&	0.397	&	…	&	CARM	\\	
J10354+694	&	M3.5\,V	&	3418	&	4.98	&	--0.01	&	< 3	&	0.388	&	…	&	CARM	co-add	\\
J10396--069	&	M2.5\,V	&	3524	&	4.91	&	--0.06	&	< 3	&	0.541	&	…	&	CARM	\\	
J10416+376	&	M4.5\,V	&	3263	&	5.07	&	0.05	&	4.1$^{j}$	&	0.212	&	…	&	CARM	\\	
J10508+068	&	M4.0\,V	&	3335	&	5.03	&	0.03	&	< 3	&	0.281	&	…	&	CARM	co-add	\\
J10520+139	&	M3.5\,V	&	3372	&	5.02	&	--0.03	&	< 3	&	0.289	&	…	&	FEROS	\\	
J11000+228	&	M2.5\,V	&	3500	&	4.94	&	--0.10	&	2.5$^{b}$	&	0.423	&	…	&	CARM	co-add	\\
J11026+219	&	M1.0\,V	&	3896	&	4.69	&	0.04	&	4.5	&	0.603	&	yes	&	CARM	co-add	\\
J11033+359	&	M1.5\,V	&	3598	&	4.87	&	--0.09	&	< 3	&	0.452	&	…	&	CARM	co-add	\\
J11054+435	&	M1.0\,V	&	3636	&	4.91	&	--0.29	&	< 3	&	0.430	&	…	&	CARM	co-add	\\
J11110+304	&	M2.0\,V	&	3753	&	4.70	&	0.14	&	< 3	&	0.599	&	…	&	CARM	co-add	\\
J11126+189	&	M1.5\,V	&	3752	&	4.73	&	0.06	&	< 3	&	0.565	&	…	&	CARM	co-add	\\
J11201--104	&	M2.0\,V	&	3540	&	4.97	&	--0.27	&	< 3	&	0.515	&	yes	&	CARM	\\	
J11289+101	&	M3.5\,V	&	3364	&	5.02	&	0.00	&	< 3	&	0.363	&	…	&	CARM	\\	
J11306--080	&	M3.5\,V	&	3419	&	4.98	&	--0.01	&	< 3	&	0.390	&	…	&	CARM	\\	
J11417+427	&	M4.0\,V	&	3358	&	4.99	&	0.13	&	< 3	&	0.381	&	…	&	CARM	co-add	\\
J11421+267	&	M2.5\,V	&	3512	&	4.90	&	--0.02	&	< 3	&	0.485	&	…	&	CARM	co-add	\\
J11467--140	&	M3.0\,V	&	3523	&	4.87	&	0.06	&	< 3	&	0.570	&	…	&	CARM	\\	
J11476+786	&	M3.5\,V	&	3359	&	5.02	&	0.00	&	< 3	&	0.258	&	…	&	CARM	co-add	\\
J11477+008	&	M4.0\,V	&	3251	&	5.10	&	--0.04	&	< 3	&	0.172	&	…	&	CARM	co-add	\\
J11509+483	&	M4.5\,V	&	3211	&	5.11	&	0.00	&	< 3	&	0.168	&	…	&	CARM	co-add	\\
J11511+352	&	M1.5\,V	&	3633	&	4.88	&	--0.18	&	< 3	&	0.506	&	…	&	CARM	co-add	\\
J11532--073	&	M2.5\,V	&	3555	&	4.87	&	0.00	&	< 3	&	0.498	&	…	&	FEROS	\\	
J12016--122	&	M3.0\,V	&	3509	&	4.88	&	0.04	&	< 3	&	0.386	&	…	&	FEROS	\\	
J12054+695	&	M4.0\,V	&	3325	&	5.02	&	0.09	&	< 3	&	0.293	&	…	&	CARM	co-add	\\
J12100--150	&	M3.5\,V	&	3365	&	4.99	&	0.12	&	< 3	&	0.433	&	…	&	CARM	co-add	\\
J12111--199	&	M3.0\,V	&	3448	&	4.97	&	--0.06	&	3.0$^{d}$	&	0.391	&	…	&	CARM	\\	
J12123+544S	&	M0.0\,V	&	3923	&	4.70	&	--0.01	&	3.9	&	0.635	&	…	&	CARM	co-add	\\
J12230+640	&	M3.0\,V	&	3528	&	4.87	&	0.03	&	< 3	&	0.529	&	…	&	CARM	co-add	\\
J12248--182	&	M2.0\,V	&	3476	&	4.98	&	--0.18	&	< 3	&	0.271	&	…	&	CARM	\\	
J12312+086	&	M0.5\,V	&	3913	&	4.71	&	--0.05	&	< 3	&	0.611	&	…	&	CARM	co-add	\\
J12350+098	&	M2.5\,V	&	3578	&	4.85	&	0.00	&	< 3	&	0.524	&	…	&	CARM	\\	
J12388+116	&	M3.0\,V	&	3429	&	4.96	&	0.04	&	< 3	&	0.513	&	…	&	CARM	co-add	\\
J12428+418	&	M4.0\,V	&	3321	&	5.07	&	--0.10	&	3	&	0.289	&	…	&	CARM	co-add	\\
J12479+097	&	M3.5\,V	&	3384	&	5.00	&	0.06	&	< 3	&	0.354	&	…	&	CARM	co-add	\\
J13196+333	&	M1.5\,V	&	3801	&	4.67	&	0.18	&	< 3	&	0.606	&	…	&	CARM	co-add	\\
J13209+342	&	M1.0\,V	&	3732	&	4.76	&	--0.01	&	< 3	&	0.576	&	…	&	CARM	co-add	\\
J13229+244	&	M4.0\,V	&	3318	&	5.05	&	0.02	&	< 3	&	0.264	&	…	&	CARM	co-add	\\
J13293+114	&	M3.5\,V	&	3431	&	4.96	&	0.04	&	< 3	&	0.394	&	…	&	CARM	\\	
J13299+102	&	M0.5\,V	&	3704	&	4.82	&	--0.15	&	< 3	&	0.562	&	…	&	CARM	co-add	\\
J13343+046	&	M0.0\,V	&	4131	&	4.60	&	0.24	&	4	&	0.723	&	…	&	FEROS	\\	
J13427+332	&	M3.5\,V	&	3359	&	5.03	&	--0.01	&	4.0$^{d}$	&	0.285	&	…	&	CARM	co-add	\\
J13450+176	&	M1.0\,V	&	3806	&	4.85	&	--0.42	&	2.0$^{k}$	&	0.572	&	…	&	CARM	co-add	\\
J13457+148	&	M1.5\,V	&	3677	&	4.79	&	--0.04	&	< 3	&	0.539	&	…	&	CARM	co-add	\\
J13458--179	&	M3.5\,V	&	3399	&	4.99	&	0.04	&	< 3	&	0.333	&	…	&	CARM	\\	
J13526+144	&	M2.0\,V	&	3670	&	4.74	&	0.13	&	< 3	&	0.519	&	…	&	FEROS	\\	
J14010--026	&	M1.0\,V	&	3719	&	4.77	&	--0.03	&	< 3	&	0.552	&	…	&	CARM	co-add	\\
J14082+805	&	M1.0\,V	&	3835	&	4.67	&	0.17	&	< 3	&	0.618	&	…	&	CARM	co-add	\\
J14152+450	&	M3.0\,V	&	3456	&	4.94	&	0.00	&	< 3	&	0.463	&	…	&	CARM	co-add	\\
J14251+518	&	M2.5\,V	&	3512	&	4.92	&	--0.08	&	< 3	&	0.449	&	…	&	CARM	co-add	\\
J14257+236E	&	M0.5\,V	&	3943	&	4.65	&	0.16	&	< 3	&	0.653	&	…	&	CARM	co-add	\\
J14257+236W	&	M0.0\,V	&	4021	&	4.63	&	0.18	&	< 3	&	0.678	&	…	&	CARM	co-add	\\
J14283+053	&	M3.0\,V	&	3455	&	4.96	&	--0.03	&	< 3	&	0.398	&	…	&	FEROS	\\	
J14294+155	&	M2.0\,V	&	3633	&	4.81	&	0.00	&	< 3	&	0.555	&	…	&	CARM	co-add	\\
J14307--086	&	M0.5\,V	&	4084	&	4.63	&	0.13	&	< 3	&	0.739	&	…	&	CARM	co-add	\\
J14342--125	&	M4.0\,V	&	3325	&	5.02	&	0.11	&	< 3	&	0.303	&	…	&	CARM	co-add	\\
J14524+123	&	M2.0\,V	&	3560	&	4.88	&	--0.05	&	< 3	&	0.516	&	…	&	CARM	co-add	\\
J14544+355	&	M3.5\,V	&	3375	&	5.00	&	0.03	&	< 3	&	0.474	&	…	&	CARM	co-add	\\
J15013+055	&	M3.0\,V	&	3413	&	5.00	&	--0.04	&	< 3	&	0.400	&	…	&	CARM	\\	
J15095+031	&	M3.0\,V	&	3480	&	4.93	&	--0.01	&	< 3	&	0.482	&	…	&	CARM	co-add	\\
J15194--077	&	M3.0\,V	&	3430	&	5.00	&	--0.09	&	< 3	&	0.330	&	…	&	CARM	co-add	\\
J15412+759	&	M3.0\,V	&	3430	&	5.02	&	--0.18	&	< 3	&	0.339	&	…	&	CARM	co-add	\\
J15474--108	&	M2.0\,V	&	3515	&	4.96	&	--0.21	&	< 3	&	0.523	&	…	&	CARM	\\	
J15598--082	&	M1.0\,V	&	3644	&	4.86	&	--0.15	&	< 3	&	0.560	&	…	&	CARM	co-add	\\
J16028+205	&	M4.0\,V	&	3310	&	5.05	&	0.02	&	< 3	&	0.249	&	…	&	CARM	co-add	\\
J16092+093	&	M3.0\,V	&	3455	&	4.98	&	--0.09	&	< 3	&	0.390	&	…	&	CARM	co-add	\\
J16120+033	&	M2.0\,V	&	3592	&	4.91	&	--0.18	&	< 3	&	0.518	&	…	&	FEROS	\\	
J16167+672N	&	M3.0\,V	&	3504	&	4.91	&	0.00	&	< 3	&	0.510	&	…	&	CARM	co-add	\\
J16167+672S	&	M0.0\,V	&	4091	&	4.62	&	0.16	&	< 3	&	0.699	&	…	&	CARM	co-add	\\
J16254+543	&	M1.5\,V	&	3516	&	4.98	&	--0.27	&	< 3	&	0.350	&	…	&	CARM	co-add	\\
J16303--126	&	M3.5\,V	&	3378	&	5.01	&	0.01	&	< 3	&	0.323	&	…	&	CARM	co-add	\\
J16327+126	&	M3.0\,V	&	3486	&	4.92	&	0.00	&	< 3	&	0.390	&	…	&	CARM	co-add	\\
J16462+164	&	M2.5\,V	&	3505	&	4.92	&	--0.05	&	< 3	&	0.484	&	…	&	CARM	co-add	\\
J16487--157	&	M1.0\,V	&	3805	&	4.68	&	0.16	&	< 3	&	0.584	&	…	&	FEROS	\\	
J16554--083N	&	M3.5\,V	&	3343	&	5.05	&	--0.04	&	2.7$^{m}$	&	0.198	&	…	&	CARM	co-add	\\
J16578+473	&	M1.5\,V	&	4300	&	4.68	&	--0.43	&	< 3	&	0.705	&	…	&	CARM	co-add	\\
J16581+257	&	M1.0\,V	&	3734	&	4.78	&	--0.08	&	< 3	&	0.572	&	…	&	CARM	co-add	\\
J16591+209	&	M3.5\,V	&	3364	&	5.06	&	--0.15	&	5.70	&	0.318	&	yes	&	FEROS	\\	
J17033+514	&	M4.5\,V	&	3237	&	5.08	&	0.06	&	< 3	&	0.171	&	…	&	CARM	co-add	\\
J17052--050	&	M1.5\,V	&	3631	&	4.82	&	--0.01	&	< 3	&	0.526	&	…	&	CARM	co-add	\\
J17071+215	&	M3.0\,V	&	3482	&	4.94	&	--0.05	&	< 3	&	0.417	&	…	&	CARM	co-add	\\
J17115+384	&	M3.5\,V	&	3415	&	4.99	&	--0.01	&	< 3	&	0.417	&	…	&	CARM	co-add	\\
J17160+110	&	M1.0\,V	&	3801	&	4.69	&	0.13	&	< 3	&	0.570	&	…	&	FEROS	\\	
J17166+080	&	M2.0\,V	&	3544	&	4.91	&	--0.10	&	< 3	&	0.449	&	…	&	CARM	co-add	\\
J17198+417	&	M2.5\,V	&	3499	&	4.93	&	--0.08	&	< 3	&	0.409	&	…	&	CARM	co-add	\\
J17303+055	&	M0.0\,V	&	3804	&	4.77	&	--0.14	&	3.3	&	0.590	&	…	&	CARM	co-add	\\
J17355+616	&	M0.5\,V	&	3874	&	4.69	&	0.06	&	3.2	&	0.606	&	yes	&	CARM	co-add	\\
J17378+185	&	M1.0\,V	&	3654	&	4.88	&	--0.22	&	3	&	0.489	&	…	&	CARM	co-add	\\
J17530+169	&	M3.0\,V	&	3392	&	5.02	&	--0.08	&	< 3	&	0.388	&	…	&	FEROS	\\	
J17578+046	&	M3.5\,V	&	3278	&	5.10	&	--0.12	&	< 3	&	0.155	&	…	&	CARM	co-add	\\
J17578+465	&	M2.5\,V	&	3459	&	4.94	&	0.00	&	< 3	&	0.447	&	…	&	CARM	co-add	\\
J18051--030	&	M1.0\,V	&	3664	&	4.87	&	--0.21	&	1.6$^{c}$	&	0.521	&	…	&	CARM	co-add	\\
J18163+015	&	M3.0\,V	&	3429	&	5.00	&	--0.10	&	< 3	&	0.346	&	…	&	FEROS	\\	
J18174+483	&	M2.0\,V	&	3515	&	4.96	&	--0.18	&	< 3	&	0.510	&	yes	&	CARM	co-add	\\
J18180+387E	&	M3.0\,V	&	3434	&	4.99	&	--0.06	&	< 3	&	0.295	&	…	&	CARM	co-add	\\
J18198--019	&	K7.0\,V	&	4133	&	4.66	&	--0.08	&	< 3	&	-	&	…	&	CARM	\\	
J18221+063	&	M4.0\,V	&	3405	&	5.00	&	0.00	&	< 3	&	0.260	&	…	&	CARM	co-add	\\
J18224+620	&	M4.0\,V	&	3227	&	5.10	&	--0.01	&	2.3$^{m}$	&	0.159	&	…	&	CARM	co-add	\\
J18240+016	&	M2.0\,V	&	3514	&	4.93	&	--0.11	&	< 3	&	0.508	&	…	&	FEROS	\\	
J18312+068	&	M1.0\,V	&	3804	&	4.71	&	0.06	&	< 3	&	0.593	&	…	&	FEROS	\\	
J18319+406	&	M3.5\,V	&	3423	&	4.99	&	--0.05	&	< 3	&	0.325	&	…	&	CARM	co-add	\\
J18346+401	&	M3.5\,V	&	3392	&	4.98	&	0.09	&	2.5$^{b}$	&	0.192	&	…	&	CARM	co-add	\\
J18353+457	&	M0.5\,V	&	3915	&	4.69	&	0.05	&	1.0$^{n}$	&	0.631	&	…	&	CARM	co-add	\\
J18363+136	&	M4.0\,V	&	3301	&	5.07	&	--0.05	&	< 3	&	0.266	&	…	&	CARM	co-add	\\
J18409--133	&	M1.0\,V	&	3788	&	4.72	&	0.06	&	3.0$^{c}$	&	0.583	&	…	&	CARM	co-add	\\
J18419+318	&	M3.0\,V	&	3473	&	4.95	&	--0.06	&	2.5$^{b}$	&	0.411	&	…	&	CARM	co-add	\\
J18427+139	&	M4.0\,V	&	3254	&	5.11	&	--0.11	&	< 3	&	0.251	&	yes	&	FEROS	\\	
J18480--145	&	M2.5\,V	&	3500	&	4.94	&	--0.09	&	< 3	&	0.453	&	…	&	CARM	co-add	\\
J18518+165	&	M0.0\,V	&	3884	&	4.71	&	--0.02	&	< 3	&	0.598	&	…	&	FEROS	\\	
J18580+059	&	M0.5\,V	&	3913	&	4.68	&	0.08	&	< 3	&	0.622	&	…	&	CARM	co-add	\\
J19032+034	&	M3.0\,V	&	3473	&	4.95	&	--0.07	&	< 3	&	0.389	&	…	&	FEROS	\\	
J19070+208	&	M2.0\,V	&	3532	&	4.95	&	--0.21	&	< 3	&	0.330	&	…	&	CARM	co-add	\\
J19072+208	&	M2.0\,V	&	3535	&	4.94	&	--0.20	&	< 3	&	0.331	&	…	&	CARM	co-add	\\
J19084+322	&	M3.0\,V	&	3439	&	4.97	&	--0.04	&	< 3	&	0.389	&	…	&	CARM	co-add	\\
J19098+176	&	M4.5\,V	&	3240	&	5.08	&	0.06	&	< 3	&	0.190	&	…	&	CARM	co-add	\\
J19169+051N	&	M2.5\,V	&	3557	&	4.86	&	0.00	&	< 3	&	0.526	&	…	&	CARM	co-add	\\
J19216+208	&	M4.5\,V	&	3249	&	5.09	&	0.02	&	3.5	&	0.187	&	…	&	CARM	co-add	\\
J19220+070	&	M3.0\,V	&	3369	&	5.05	&	--0.16	&	< 3	&	0.221	&	…	&	FEROS	\\	
J19251+283	&	M3.0\,V	&	3405	&	5.00	&	0.00	&	< 3	&	0.398	&	…	&	CARM	co-add	\\
J19346+045	&	M0.0\,V	&	4054	&	4.69	&	--0.20	&	3.3	&	0.632	&	…	&	CARM	co-add	\\
J20011+002	&	M2.0\,V	&	3562	&	4.91	&	--0.13	&	< 3	&	0.525	&	…	&	FEROS	\\	
J20187+158	&	M2.5\,V	&	3514	&	4.91	&	--0.04	&	< 3	&	0.449	&	…	&	FEROS	\\	
J20305+654	&	M2.5\,V	&	3475	&	4.96	&	--0.08	&	< 3	&	0.415	&	…	&	CARM	co-add	\\
J20336+617	&	M4.0\,V	&	3368	&	4.98	&	0.14	&	< 3	&	0.420	&	…	&	CARM	co-add	\\
J20405+154	&	M4.5\,V	&	3236	&	5.09	&	0.03	&	< 3	&	0.189	&	…	&	CARM	co-add	\\
J20407+199	&	M2.5\,V	&	3475	&	4.96	&	--0.07	&	< 3	&	0.528	&	…	&	FEROS	\\	
J20450+444	&	M1.5\,V	&	3591	&	4.89	&	--0.14	&	< 3	&	0.480	&	…	&	CARM	co-add	\\
J20525--169	&	M4.0\,V	&	3313	&	5.06	&	--0.01	&	< 3	&	0.242	&	…	&	CARM	co-add	\\
J20533+621	&	M0.5\,V	&	3828	&	4.71	&	0.03	&	< 3	&	0.597	&	…	&	CARM	co-add	\\
J20556--140N	&	M4.0\,V	&	3372	&	5.02	&	0.01	&	< 3	&	0.334	&	…	&	CARM	co-add	\\
J20567--104	&	M2.5\,V	&	3523	&	4.89	&	0.00	&	< 3	&	0.502	&	…	&	CARM	co-add	\\
J21019--063	&	M2.5\,V	&	3521	&	4.90	&	--0.05	&	< 3	&	0.513	&	…	&	CARM	co-add	\\
J21057+502	&	M3.5\,V	&	3543	&	4.83	&	0.14	&	< 3	&	0.317	&	…	&	HRS	\\	
J21127--073	&	M3.5\,V	&	3471	&	4.96	&	--0.07	&	< 3	&	0.328	&	…	&	HRS	\\	
J21152+257	&	M3.0\,V	&	3657	&	4.69	&	0.28	&	< 3	&	0.397	&	…	&	CARM	co-add	\\
J21164+025	&	M3.0\,V	&	3475	&	4.95	&	--0.05	&	< 3	&	0.402	&	…	&	CARM	co-add	\\
J21221+229	&	M1.0\,V	&	3705	&	4.83	&	--0.19	&	3.7	&	0.590	&	…	&	CARM	co-add	\\
J21348+515	&	M3.0\,V	&	3484	&	4.92	&	0.00	&	< 3	&	0.494	&	…	&	CARM	co-add	\\
J21463+382	&	M4.0\,V	&	3304	&	5.06	&	--0.01	&	< 3	&	0.168	&	…	&	CARM	co-add	\\
J21466--001	&	M4.0\,V	&	3346	&	5.02	&	0.05	&	4.0$^{d}$	&	0.292	&	…	&	CARM	co-add	\\
J21466+668	&	M4.0\,V	&	3355	&	5.01	&	0.05	&	< 3	&	0.258	&	…	&	CARM	co-add	\\
J21472--047	&	M4.5\,V	&	3273	&	5.10	&	--0.11	&	< 3	&	0.201	&	…	&	HRS	\\	
J21574+081	&	M1.5\,V	&	3858	&	4.62	&	0.34	&	< 3	&	0.596	&	…	&	FEROS	\\	
J22020--194	&	M3.5\,V	&	3431	&	4.97	&	--0.01	&	< 3	&	0.362	&	…	&	CARM	co-add	\\
J22021+014	&	M0.5\,V	&	3914	&	4.69	&	0.05	&	< 3	&	0.600	&	…	&	CARM	co-add	\\
J22057+656	&	M1.5\,V	&	3653	&	4.85	&	--0.15	&	3.9	&	0.314	&	…	&	CARM	co-add	\\
J22096--046	&	M3.5\,V	&	3454	&	4.96	&	--0.01	&	< 3	&	0.531	&	…	&	CARM	co-add	\\
J22115+184	&	M2.0\,V	&	3554	&	4.90	&	--0.10	&	< 3	&	0.580	&	…	&	CARM	co-add	\\
J22125+085	&	M3.0\,V	&	3500	&	4.92	&	--0.04	&	< 3	&	0.381	&	…	&	CARM	co-add	\\
J22231--176	&	M4.5\,V	&	3196	&	5.12	&	--0.05	&	< 3	&	0.173	&	…	&	CARM	co-add	\\
J22252+594	&	M4.0\,V	&	3383	&	5.00	&	0.05	&	< 3	&	0.385	&	…	&	CARM	co-add	\\
J22298+414	&	M4.0\,V	&	3318	&	5.03	&	0.10	&	< 3	&	0.254	&	…	&	CARM	co-add	\\
J22330+093	&	M1.0\,V	&	3660	&	4.87	&	--0.22	&	2.64$^{c}$	&	0.525	&	…	&	CARM	co-add	\\
J22503--070	&	M0.5\,V	&	3895	&	4.73	&	--0.10	&	< 3	&	0.600	&	…	&	CARM	co-add	\\
J22532--142	&	M4.0\,V	&	3359	&	5.01	&	0.06	&	2.5$^{b}$	&	0.370	&	…	&	CARM	co-add	\\
J22559+178	&	M1.0\,V	&	3824	&	4.70	&	0.07	&	< 3	&	0.599	&	…	&	CARM	co-add	\\
J22565+165	&	M1.5\,V	&	3787	&	4.70	&	0.10	&	2.5$^{b}$	&	0.601	&	…	&	CARM	co-add	\\
J23113+085	&	M3.5\,V	&	3404	&	4.99	&	0.01	&	< 3	&	0.330	&	…	&	CARM	co-add	\\
J23175+063	&	M3.0\,V	&	3481	&	4.93	&	--0.01	&	< 3	&	0.400	&	…	&	FEROS	\\	
J23216+172	&	M4.0\,V	&	3361	&	4.99	&	0.14	&	< 3	&	0.437	&	…	&	CARM	co-add	\\
J23234+155	&	M2.0\,V	&	3635	&	4.81	&	0.00	&	< 3	&	0.509	&	…	&	FEROS	\\	
J23245+578	&	M1.0\,V	&	3824	&	4.69	&	0.12	&	0.5$^{c}$	&	0.606	&	…	&	CARM	co-add	\\
J23340+001	&	M2.5\,V	&	3553	&	4.87	&	0.00	&	< 3	&	0.476	&	…	&	CARM	co-add	\\
J23381--162	&	M2.0\,V	&	3545	&	4.92	&	--0.13	&	< 3	&	0.508	&	…	&	CARM	co-add	\\
J23419+441	&	M5.0\,V	&	3144	&	5.13	&	0.05	&	1.2$^{j}$	&	0.141	&	…	&	CARM	co-add	\\
J23431+365	&	M4.0\,V	&	3247	&	5.10	&	--0.05	&	2.6$^{m}$	&	0.208	&	…	&	CARM	co-add	\\
J23492+024	&	M1.0\,V	&	3657	&	4.84	&	--0.12	&	< 3	&	0.465	&	…	&	CARM	co-add	\\
J23556--061	&	M2.5\,V	&	3648	&	4.76	&	0.11	&	< 3	&	0.598	&	…	&	CARM	co-add	\\
J23577+233	&	M3.5\,V	&	3419	&	4.98	&	0.00	&	5.20	&	0.423	&	…	&	FEROS	\\	
J23585+076	&	M3.0\,V	&	3470	&	4.94	&	0.00	&	< 3	&	0.507	&	…	&	CARM	co-add	\\
\end{longtable}

$^{a}$Carmencita identifier (Karmn), spectral type, effective temperature, surface gravity, metallicity, $v \sin i$, mass, Ca~{\sc ii} emission flag, 
and instrument with which the spectrum was obtained (CARMENES --``CARM co-add'' for co-added, ``CARM'' for single spectra), CAFE, FEROS, HRS). Rotational velocities ($v \sin i$) 
from \cite{Jeffers2018}, if no other reference is given\newline
$^{b}$ \cite{Browning2010}, $^{c}$ \cite{Houdebine2010}, $^{d}$ \cite{Reiners2012}, $^{e}$ \cite{Martinez2010}, $^{f}$ \cite{ReinersBasri2007}, $^{g}$ \cite{Antonova2013}, 
$^{h}$ \cite{StaufferHartmann1986}, $^{j}$ \cite{Jenkins2009}, $^{k}$ \cite{GlebockiGnacinski2005}, $^{m}$ \cite{MohantyBasri2003}, $^{n}$ \cite{MarcyChen1992}.


\end{document}